\documentclass[12pt,preprint]{aastex}

\usepackage{graphicx}
\usepackage{subfig}
\usepackage{epstopdf}
\usepackage{verbatim}
\usepackage{multirow}
\usepackage{float}
\usepackage{graphics,epsfig}
\usepackage{natbib}
\usepackage{lscape}

\def\spose#1{\hbox to 0pt{#1\hss}}
\def\lta{\mathrel{\spose{\lower 3pt\hbox{$\mathchar"218$}}
     \raise 2.0pt\hbox{$\mathchar"13C$}}}
\def\gta{\mathrel{\spose{\lower 3pt\hbox{$\mathchar"218$}}
     \raise 2.0pt\hbox{$\mathchar"13E$}}}

\def\figure#1#2 {\par{\narrower\noindent {\bf Fig. #1}
   \hskip 2mm #2\par}\bigskip\noindent}
\def\table#1#2 {\par{\narrower\noindent {\bf Tab. #1}
   \hskip 2mm #2\par}\bigskip\noindent}

\def\registered{{\ooalign{\hfil\raise .00ex\hbox{\scriptsize R}\hfil\crcr\mathhexbox20D}}}

\usepackage{amsmath}
\usepackage{accents}
\newlength{\dhatheight}

\slugcomment{Version: \enskip \today}

\shorttitle{Fitting Formulae for Binary Habitable Zones}

\shortauthors{Zh. Wang \& M. Cuntz}

\begin{document}


\title{
Fitting Formulae and Constraints for the Existence of \\
S-type and P-type Habitable Zones \\
in Binary Systems
}

\bigskip
\bigskip

\author{Zhaopeng Wang and Manfred Cuntz}

\bigskip

\affil{Department of Physics, University of Texas at Arlington, \\
Arlington, TX 76019, USA}
\email{zhaopeng.wang@mavs.uta.edu; cuntz@uta.edu}

\bigskip

%%%%%%%%%%%%%%%%%%%%%%%%%%%%%%%%%%%%%%%%%%%%%%%%%%%%%%%%%%%%%%%%%%%%%%%%

\begin{abstract}
We derive fitting formulae for the quick determination of the existence of S-type and P-type
habitable zones in binary systems.  Based on previous work, we consider the limits of the
climatological habitable zone in binary systems (which sensitively depend on the system parameters)
based on a joint constraint encompassing planetary orbital stability and a habitable region for
a possible system planet.  Additionally, we employ updated results on planetary climate models
obtained by Kopparapu and collaborators.  Our results are applied to four P-type systems
(Kepler-34, Kepler-35, Kepler-413, and Kepler-1647) and two S-type systems (TrES-2 and KOI-1257).
Our method allows to gauge the existence of climatological habitable zones for these systems
in a straightforward manner with detailed consideration of the observational uncertainties.
Further applications may include studies of other existing systems as well as systems to be
identified through future observational campaigns.
\end{abstract}


\clearpage

%%%%%%%%%%%%%%%%%%%%%%%%%%%%%%%%%%%%%%%%%%%%%%%%%%%%%%%%%%%%%%%%%%%%%%%%

\section{Introduction}

Several decades of detailed observations revealed that stellar binary
systems constitute a notable component of our Galactic neighborhood
\citep[e.g.,][]{duq91,pat02,egg04,rag06,rag10,roe12}.
An important aspect of this type of research is the plethora of
discoveries of planets in many of those systems.  Generally there are
two types of possible planetary orbits \citep[e.g.,][]{dvo82}: planets
orbiting one of the binary components are said to be in S-type orbits,
while planets orbiting both binary components are said to be in P-type
orbits.  In fact, since 1989, 83 planet-hosting binary systems, encompassing
63 planets in S-type obits and 20 planets in P-type orbits, have been detected,
mostly based on the radial velocity method and transit method.  A survey
about exoplanetary systems of binary stars with stellar separations less
than 100~au was given by \cite{baz17}; it also considers the effects of
secular resonances on the systems' habitability.

In 1989, HD~114762b in the constellation Coma Berenices has tentatively
been identified as an exoplanet, thus being the first possible planet
around a main-sequence star other than the Sun and, incidently, the first
possible planet located in a binary system.  In 2012, this planet was
finally confirmed based on the radial velocity method.  A more recent
example is HD~87646b, a planet in a close binary system with a 22~au
separation distance \citep{ma16}.  This system contains two substellar objects in
S-type orbits, which makes it the first close binary system known to host
more than one substellar companion.  Other examples of planets in binary
systems include Kepler-413b \citep{kos14} and Kepler-453b \citep{wel15}.
Both Kepler-413b and Kepler-453b are in P-type orbits, also called
circumbinary orbits.  However, S-type orbits are much more frequent,
and in some systems the planets are in orbit around quasi-single stars.
For example,  Kepler-432b, a hot Jupiter-type planet orbits a giant star
that is part of a super-wide binary system with a separation distance
of 750~au \citep{ort15}.  Some of the S-type and P-type orbits are 
located within the stellar habitable zones (HZs).
These systems often receive special attention as they
inspire detailed studies about the planet's long-term orbital stability
and its potential for hosting exolife.

In previous studies, focusing on habitable zones in stellar binary systems,
presented by \citep{cun14,cun15} denoted as Paper I and II, respectively,
henceforth, a joint constraint of radiative habitable zones (RHZs, based on
stellar radiation) and orbital stability was considered.  Previous results were
given by \cite{egg12,egg13}, \cite{kan13}, \cite{kal13}, \cite{hag13}, among
others\footnote{We wish to draw the reader's attention to the online calculator
{\tt BinHab} \citep{cunb14}, hosted at The University of Texas at Arlington (UTA),
which allows the calculation of habitable regions in binary systems based on the
developed method.  Another online calculator with similar capacities has been
given by \cite{mul14}.  \cite{zul16} pursued a comparison study between these tools
and found that their results are consistent with each other.}.
Paper~II also takes into account the eccentricity of binary components, which is found
to adversely affect the width of the HZs.  RHZs, encompassing the conservative, general and
recent Venus / early Mars HZs (henceforth referred to as CHZ, GHZ, and RVEM, respectively),
are defined in accordance to the respective limits identified in the Solar System.
Our work also takes into account detailed results obtained by \cite{kop13,kop14}.
This work employs updated 1D radiative--convective, cloud-free climate model, which
among other improvements are based on revised H$_2$O and CO$_2$
absorption coefficients.  Previous results about limits of stellar habitable zones
have been given by, e.g., \cite{kas93} and \cite{und03}.  The latter explores how
HZs are impacted by stellar evolution. 

Our paper is structured as follows. In Section~2, we describe the theoretical
approach, including general background information.  The fitting procedure is
outlined in Section~3.  Section~4 offers applications to observed systems,
encompassing systems with S-type and P-type planets.
Our summary and conclusions are given in Section~4.

%%%%%%%%%%%%%%%%%%%%%%%%%%%%%%%%%%%%%%%%%%%%%%%%%%%%%%%%%%%%%%%%%%%%%%%%

\section{Methodology}

\subsection{Theoretical Background}

Based on the radiative energy fluxes received by system planets from the two binary components,
the habitable limits could, in principle, be defined similarly to those within the Solar System,
amounting to the concept of the RHZs; see Section~1.  Following previous
work\footnote{This subsection is merely intended as supplementary information; it summarizes
materials previously given in Paper I and II.} the RHZs can be calculated based on

\begin{equation}
	\setlength{\abovedisplayskip}{5pt}
	\setlength{\belowdisplayskip}{5pt}
	\frac{L_{1}}{S_{{\rm rel},1l}d^{2}_{1}} +  \frac{L_{2}}{S_{{\rm rel},2l}d^{2}_{2}}= \frac{L_{\odot}}{s^{2}_{l}}
\end{equation}
with $d_{1}$ and $d_{2}$ denoting the distances from to the binary components (see Fig.~1),
$L_{1}$ and $L_{2}$ indicating the stellar luminosities, and $s_{l}$ standing for one of
the solar habitability limits (see Table~1). $S_{{\rm rel,}{il}}$ with $i=1,2$ is the stellar flux
in units of solar constant, which depends on the effective temperature of binary stars.
Since $d_{1}$ and $d_{2}$ can be represented by a function of $z$, the distance from the center
of binary system, a quartic equation for $z$, can be obtained after algebraic transformations. 

Hence, the RHZ, an annulus around each star (S-type) or both stars (P-type), is thus given as
\begin{equation}
	\setlength{\abovedisplayskip}{5pt}
	\setlength{\belowdisplayskip}{5pt}
	\textrm{RHZ(z)} = \textrm{Min}({\cal R}(z,\varphi))|s_{l,{\rm out}} - \textrm{Max}({\cal R}(z,\varphi))|s_{l,{\rm in}}
\end{equation}
Here ${\cal R}(z,\varphi)$ describe the borders of the RHZs, with $z$ and $\varphi$ denoting the
polar coordinates.  Additionally, $s_{l,{\rm in}}$ and $s_{l,{\rm out}}$ describe the parameters
tagging respectively the inner and outer limits of the stellar RHZ; see Table~1.

If a planet is assumed to stay in the HZ for timespan of astrobiological significance,
a stable orbit is required.  Using the fitting equations developed by \cite{hol99},
the planetary orbital stability limits\footnote{The formulae
of orbital stability by \cite{hol99} are based on $10^4$ binary periods, a time scale
significantly shorter than required for the installment of astrobiology.  However, more
recent studies by \cite{pil02} for S-type systems based on the Fast Lyapunov Integrator
indicate that the orbital stability limits of \cite{hol99} are also valid for notably
longer time scales such as $10^6$ binary periods especially for systems with planets
in nearly circular orbits.  Nevertheless, improvements of the \citeauthor{hol99}
formulae for general systems for long time scales, ideally encompassing billions
of years, should be considered a topic of high priority due to their significance
for future astrobiological studies.} are obtained.  They convey an upper limit
as the distance from the stellar primary for S-type orbits, and a lower limit
measured from the mass center of the binary system for P-type orbits.  Additionally,
following the terminology of Paper~I, ST-type and PT-type HZs denote the cases
when the widths of the HZs are impacted by the orbital stability limits and,
therefore, the corresponding RHZs are truncated.  Consequently, the width of the
P/PT-type HZ (if existing) is given by
\begin{equation}
{\rm Width}~(P/PT) \ = \ {\rm RHZ}_{\rm out} - {\rm Max} \big( {\rm RHZ}_{\rm in} , a_{\rm cr} \big) \ ,
\end{equation}
and the the width of the S/ST-type HZ (if existing) is given by
\begin{equation}
{\rm Width}~(S/ST) \ = \  {\rm Min} \big( {\rm RHZ}_{\rm out} , a_{\rm cr} \big) - {\rm RHZ}_{\rm in} \ .
\end{equation}

Here ${\rm RHZ}_{\rm in}$ and ${\rm RHZ}_{\rm out}$ denote the inner and outer limits
of the RHZs, respectively, and $a_{\rm cr}$ denotes the orbital stability limit.
Equations (3) and (4) are relevant for devising the fitting formulae for the
existence of P-type and S-type HZs, the main focus of this study.

%%%%%%%%%%%%%%%%%%%%%%%%%%%%%%%%%%%%%%%%%%%%%%%%%%%%%%%%%%%%%%%%%%%%%%%%

\subsection{General Analysis}

Various sets of binary systems, encompassing both systems of equal and non-equal masses,
have been studied to examine the existence of their HZs based on the radiative criterion,
as described by the stellar luminosities, as well as the orbital stability criterion for
system planets.  Information on the adopted stellar parameters, chosen for cases of
theoretical main-sequence stars, are given in Table~1 and 2.  Regarding the stellar HZs,
we focus on the GHZ and RVEM (see Sect. 2.1) and consider stars with masses $M_1$ and
$M_2$ of 0.50~$M_{\odot}$, 0.75~$M_{\odot}$, 1.00~$M_{\odot}$, and 1.25~$M _{\odot}$;
see Figures 2 to 5 for details.

For systems of masses $M_{1} = M_{2} = 1.0~M_{\odot}$, in case of $e_{b}$ = 0,
the semi-major axis $a_{\rm bin}$ is required to be smaller than 0.97~au for the P/PT-type GHZ to
exist and smaller than 1.03~au for the P/PT-type RVEM to exist.  Regarding S/ST-type HZs,
$a_{\rm bin}$ needs to be larger than 3.72~au and larger than 2.93~au for the GHZ and RVEM to exist,
respectively.  Larger eccentricities barely affect the existence of P/PT-type HZs; however,
they notably affect the existence of the S/ST-type HZs.  For $e_{b}$ = 0.50, $a_{\rm bin}$ is
required to be larger than 8.44~au and larger than 6.66~au for S/ST-type GHZ and
RVEM to exist, respectively.

Different values for the existence of P/PT and S/ST-type HZs are obtained for other
kinds of equal-mass systems.  For systems with masses $M_{1} = M_{2} = 0.50~M_{\odot}$,
in case of $e_{b}$ = 0, $a_{\rm bin}$ is required to be smaller than 0.22~au for the P/PT-type GHZ
to exist and smaller than 0.23~au for the P/PT-type RVEM to exist.  Regarding S/ST-type HZs,
$a_{\rm bin}$ needs to be larger than 0.76~au and larger than 0.60~au for the GHZ and RVEM to exist,
respectively.  Again, larger eccentricities barely affect the existence of P/PT-type HZs;
however, they impact the existence of the S/ST-type HZs, as expected.  For $e_{b}$ = 0.50,
$a_{\rm bin}$ is required to be larger than 1.74~au and larger than 1.37~au for S/ST-type GHZ
and RVEM to exist, respectively.

We also investigated non-equal mass systems, which generally are considered more significant
than equal-mass systems.  In systems of $M_{1} = 1.00~M_{\odot}$ and $M_{2} = 0.50~M_{\odot}$,
$a_{\rm bin}$ is required to be smaller than 0.81~au and smaller than 0.86~au for the P/PT-type GHZ
and RVEM, respectively, in case of $e_{b}$ = 0.  Furthermore, $a_{\rm bin}$ needs to be larger than
2.83~au and larger than 2.23~au to allow for the existence of the S/ST-type GHZ and RVEM,
respectively.  Again, high eccentricities barely affect the existence of P/PT-type HZs;
however, they impact the existence of the S/ST-type HZs as already discussed for equal-mass
systems.  For $e_{b}$ = 0.50, $a_{\rm bin}$ is required to be larger than 6.78~au and larger than
5.35~ au for S/ST-type GHZ and RVEM to exist, respectively.

Moreover, we also considered systems of $M_{1} = 1.25~M _{\odot}$ and $M_{2} = 0.75~M_{\odot}$.
The case of $e_{b}$ = 0 requires $a_{\rm bin}$ to be smaller than 1.20~au and smaller than 1.27~au for
the P/PT-type GHZ and RVEM, respectively.  Regarding S/ST-type HZs, $a_{\rm bin}$ is required to be
larger than 4.36~au for the GHZ and larger than 3.41~au for the RVEM regarding $e_{b}$ = 0.
Furthermore, $a_{\rm bin}$ is required to be larger than 10.17~au for the GHZ and larger than 8.03~au
for the RVEM and $e_{b}$ = 0.50.  Even higher values of $a_{\rm bin}$ are needed for both the GHZ
and RVEM in case of eccentricities beyond 0.50.

In summary, the existence of P/PT-type HZs is barely affected by the eccentricity of the
stellar system and solely controlled by $M_1$ and $M_2$ (or, say, $L_1$ and $L_2$).  However,
relatively large semi-major axes $a_{\rm bin}$ are required for the existence of S/ST-type HZs in highly
eccentric systems.  Large values of $a_{\rm bin}$ always ensure S/ST-type HZs, as in this case, the
stellar habitable environments are in essence those of single stars.  In non-equal mass systems
with $M_1 + M_2$ considered as fixed, the P/PT-type HZs are barely impacted compared to
equal-mass systems, but higher values for $a_{\rm bin}$ are mandated for the existence of S/ST-type
HZs especially for systems with high eccentricities for the binary components.

%%%%%%%%%%%%%%%%%%%%%%%%%%%%%%%%%%%%%%%%%%%%%%%%%%%%%%%%%%%%%%%%%%%%%%%%

\section{Fitting Procedure}

The main aspect of our work concerns the derivation of fitting formulae for the existence
of P/PT-type HZs and S/ST-type HZs for binary systems consisting of main-sequence stars.
Through applying the least-squared method, fitting is done in two steps: first, fitting
$a_{\rm bin}$ versus $e_{b}$ by assuming fixed masses as reference (aimed at catching the sets
of parameters where the HZs cease to exist) and, second, fitting the coefficients with
stellar masses to allow the expansion of the formulae for general binary systems. 
In the first step, the Bayesian information criterion (BIC) and mean absolute percentage 
error are taken into account. The BIC is used in the second step as well for the mass
fitting determination.

For P/PT-type cases, the $a_{\rm bin}$ versus $e_{b}$ fitting is done using a polynomial equation,
which is
\begin{equation}
	\setlength{\abovedisplayskip}{5pt}
	\setlength{\belowdisplayskip}{5pt}
a_{\rm bin} = \alpha_{0} + \alpha_{1}e_{b} + \alpha_{2}e_{b}^{2} \ .
\end{equation}
For S/ST-type cases, the $a_{\rm bin}$ versus $e_{b}$ fitting is done using a cubic equation, placed
as exponent, which reads
\begin{equation}
	\setlength{\abovedisplayskip}{5pt}
	\setlength{\belowdisplayskip}{5pt}
a_{\rm bin} = e^{\beta_{0} + \beta_{1}e_{b} + \beta_{2}e_{b}^{2} + \beta_{3}e_{b}^{3}} \ .
\end{equation}

The coefficients for selected systems are shown in Table~3. For $e_{b} = 0$, systems
with masses of $M_{1} = M_{2} = 1.00~M_{\odot}$ have 0.960~au for the
fitting results and 0.97~au regarding the data for the P/PT-type GHZ, and
1.016~au in the fitting results and 1.03~au regarding the data for the P/PT-type
RVEM to exist.  For S/ST-type HZs, the fit yields  3.597~au and 2.773~au for the
GHZ and RVEM, respectively, with data noted as 3.72~au and 2.93~au, respectively.

Keep $e_{b}$ to be zero, systems with $M_{1} = M_{2} = 0.50~M_{\odot}$, render
0.215~au, 0.230~au, 0.720~au, and 0.568~au in the fitting of P-GHZ, P-RVEM, S-GHZ,
and S-RVEM, respectively.  Conversely, the data based on the method as given in
Sect.~2.1 are given as 0.22~au, 0.23~au, 0.76~au, and 0.60~au, respectively.
In systems with $M_{1} = 1.00~M_{\odot}, M_{2} = 0.50~M_{\odot}$, the fitting results
read 0.805~au, 0.850~au, 2.702~au, and 2.109~au, respectively with  $e_{b}$ as zero
for P-GHZ, P-RVEM, S-GHZ, and S-RVEM, respectively.   Furthermore, 0.81~au, 0.86~au,
2.83~au, and 2.23~au are the values for corresponding data.
For the case of $M_{1} = 1.25~M _{\odot}$ and $M_{2} = 0.75~M_{\odot}$ with
$e_{b}$ as zero, the results for the fitting of P-GHZ, P-RVEM, S-GHZ, and S-RVEM
are given as 1.185~au, 1.253~au, 4.229~au, and 3.287~au, respectively.  Here the
data are 1.20~au, 1.27~au, 4.36~au, and 3.41~au, respectively, again showing very
close agreement.  The various fitting coefficients are listed in Table~4.

To enhance the universal applicability of the fitting formulae, we also explored
the relation between the coefficents in $a_{\rm bin}$ versus $e_{b}$ fitting, and the
stellar masses of the binary systems.  For the coefficents in the P-type equation,
$\alpha$ is represented by

\begin{equation}
	\setlength{\abovedisplayskip}{5pt}
	\setlength{\belowdisplayskip}{5pt}
\alpha_{i} = A_{i0} + A_{i1}M_{1} + A_{i2}M_{2}
\end{equation}

Furthermore, for the coefficents in the S-type equation,
$\beta$ is represented by

\begin{equation}
	\setlength{\abovedisplayskip}{5pt}
	\setlength{\belowdisplayskip}{5pt}
\beta_{i} = B_{i0} + B_{i1}M_{1} + B_{i2}M_{2} + B_{i3}M_{1}^{2} + B_{i4}M_{1}^{3}
\end{equation}

The BIC for the cases of mass fitting are listed
in Table~5.  By adding terms to the equation only containing constant and linear terms
of $M_1$ and $M_2$, the BIC varies and indicates that
P-type cases favor adding nothing, whereas S-type cases prefer adding
$M_1^2$ and $M_1^3$, as done as part of the process.  The general fitting
coefficients for P-type and S-type HZs are listed in Table~6 and 7, respectively. 

Applying the calculated coefficients from stellar masses to the $a_{\rm bin}$ versus $e_{b}$
equations, fitting results are plotted as well as the data for comparison
(see Fig.~3).  Most of the fits are virtually indistinguishable from the data.
Percent errors of the fits for selected cases are provided in Tables~8 to 10.
Cases not shown here reveal similar results.  In Table~10, the coefficients of
determination measuring the goodness of the fit, are given for reference.
The percentage errors are calculated as

\begin{equation}
	\setlength{\abovedisplayskip}{5pt}
	\setlength{\belowdisplayskip}{5pt}
	\textrm{Percentage Error} = \left| \frac{{\rm data} - {\rm fitting}}{{\rm data}} \right|
\end{equation}

In summary, through employing a two-step fitting procedure,
fully acceptable results are obtained for the fitting equations in response to the
existence of GHZ and RVEM HZs depending on the system parameters (i.e., $a_{\rm bin}$, $e_b$,
$M_1$, and $M_2$).

Using the fitting formulae given above (see Eqs. 5 to 8), several observed binary systems
have been studied in more detail to inquire on the existence of the stellar HZs,
encompassing both GHZs and RVEM HZs (see Sect.~4).  Generally, the minimum $a_{\rm bin}$ for
S-type HZs to exist would increase as either stellar mass increases.  As for the P-type cases,
the maximum $a_{\rm bin}$ is decreased with either stellar mass decreased.  Thus, the maximum masses
of binary components were considered for S-type HZs based on their errors for their existence,
and minimum masses have been taken into account for the non-existence of HZs.  In contrast,
P-type HZs consider stellar minimum masses (as defined by the respective observational
uncertainties) for their existence, and maximum masses for the non-existence of the HZs.
As for the eccentricity, the largest eccentricity (as set by observational constraints)
should be considered for the study of both S-type and P-type HZs, as it corresponds to
the most adverse outcome. 

%%%%%%%%%%%%%%%%%%%%%%%%%%%%%%%%%%%%%%%%%%%%%%%%%%%%%%%%%%%%%%%%%%%%%%%%

\section{Applications to Observed Systems}

\subsection{P-type Systems}

\subsubsection{Kepler-34}

\cite{wel12} have reported transiting circumbinary planets both regarding Kepler-34
and Kepler-35; their study also conveys detailed information about the system's data
(see Table~11).  Kepler-34 has two Sun-like stars revolving around each in
$27.7958103^{+0.0000016}_{-0.0000015}$~d, with stellar masses to be
$1.0479^{+0.0033}_{-0.0030} M_{\odot}$ and 1.0208$\pm$0.0022 $M_{\odot}$,
respectively. The system possesses a $0.220^{+0.011}_{-0.010} M_{J}$ 
circumbinary gas giant, i.e., somewhat less massive than Saturn, with a $1.0896 \pm 0.0009$~au
semi-major axis and a $0.182^{+0.016}_{-0.020}$ eccentricity.
By measuring the effective temperature and metallicity of both stars, an age between 5 and 6~Gyr
has been deduced \citep{yi01}, based on Yonsei--Yale theoretical models of stellar evolution,
and thus the stars should still be in their main-sequence stages.
The semi-major axis of the binary is $0.22882^{+0.00019}_{-0.00018}$, and the eccentricity is 
$0.52087^{+0.00052}_{-0.00055}$.  Considering the smallest possible stellar masses and 
largest possible eccentricity, both P-type GHZ and P-type RVEM HZs are expected to exist,
noting that those should require for $a_{\rm bin}$ to be less than 0.683~au and less than 0.722~au,
respectively.  These criteria are fulfilled based on the observational data (see Fig.~6).
\cite{wel12} also pointed out that the circumbinary planet is located interior to the HZ.

%(https://arxiv.org/abs/1204.3955)

\subsubsection{Kepler-35}

Kepler-35 is known to have a $0.127^{+0.020}_{-0.020} M_{J}$ circumbinary gas giant orbiting hosting starts 
on a nearly circular orbit ($e_{b}=0.042^{+0.007}_{-0.004}$); see \cite{wel12} for details.
The planet, which has a semi-major axis to be $0.17617^{+0.00029}_{-0.00030}$ au, is within the P-type HZs. 
The primary star of Kepler-35 has a mass to be $0.8877^{+0.0051}_{-0.0053} M_{\odot}$, and the secondary being 
$0.8094^{+0.0042}_{-0.0045} M_{\odot}$, with an orbital period of $20.733666^{+0.000012}_{-0.000012}$ days. 
Based on the Yonsei--Yale theoretical models of stellar evolution \citep{yi01}, the age of this system is
about 8 Gyr to 12 Gyr.  This is larger than the solar age; however, based on the masses of the two
stellar components, this system is still considered to be composed of main-sequence stars.
The binary system has a semi-separation of $0.17617^{+0.00029}_{-0.00030}$~au, and the eccentricity is
given as $0.1421^{+0.0014}_{-0.0015}$.  The system's semi-major axis is clearly less than the requirements
for P-type HZs to exist, which are 0.674~au and 0.712~au for the GHZ and RVEM (see Fig.~6).
\cite{wel12} pointed out that the circumbinary planet is located interior to the HZ.

%(https://arxiv.org/abs/1204.3955)

\subsubsection{Kepler-413}

Following \cite{kos14}, Kepler-413 has a $0.820^{+0.015}_{-0.014} M_{\odot}$ K dwarf as primary,
and a $0.5423^{+0.0081}_{-0.0073} M_{\odot}$ M dwarf as secondary (see Table~11).  The two stars
orbit each other on a nearly circular orbit, which has an eccentricity of $0.0365^{+0.0023}_{-0.0021}$.
The orbital period is given as $10.1161114^{+0.0000099}_{-0.0000101}$~d. Kepler-413b, a
$67^{+22}_{-21} M_{\oplus}$ circumbinary planet, orbiting both stars with $0.3553^{+0.0020}_{-0.0018}$~au
as semi-major axis and $0.1181^{+0.0018}_{-0.0017}$ as eccentricity, is slightly outside of the GHZ,
but insight of the RVEM.  Fitting with the minimum possible stellar masses and maximum possible
binary eccentricity (the most adverse choices for the existence of the circumstellar HZs),
the P-type GHZ and RVEM require $a_{\rm bin}$ to be smaller than 0.572~au and 0.605~au to allow their
existence.  The semi-major axis of the system is $0.10148^{+0.00057}_{-0.00052}$~au, which
satisfies the requirements for circumbinary HZs to exist (see Fig.~6).

%(https://arxiv.org/abs/1401.7275)
%(https://www.nasa.gov/ames/kepler-finds-a-very-wobbly-planet)

\subsubsection{Kepler-1647}

Following \cite{kos16}, Kepler-1647b has been identified a $483\pm206 M_{\oplus}$ gas giant
in the eclipsing binary system Kepler-1647. The semi-major axis of the planet is 2.7205$\pm$0.0070~au,
which is within the P-type GHZ.  The primary star is estimated to be a F8 star with stellar mass as
$M_1 = 1.2207\pm0.0112M_{\odot}$.  The secondary star is similar to the Sun; it has a mass of
$M_2 = 0.9678\pm0.0039M_{\odot}$ (see Table~11).  The estimated age of the system, identified
as approximately 4.4~Gyr, corresponds to mid-age main-sequence.
The binary orbital period is given as 11.2588179$\pm$0.0000013~d, with 0.1276$\pm$0.0002~au as
binary semi-major axis and 0.1602$\pm$0.0004 as binary eccentricity.  Taking the largest eccentricity
the smallest possible stellar masses, as conveyed by the observational results, as reference, it
requires $a_{\rm bin}$ to be less than 1.037~au for P-type GHZ to exist, and 1.096~au for P-type RVEM
to exist.  Fitting results for this system are shown in Figure~6, which clearly show that
both the GHZ and RVEM are able to exist in this system.

%(https://arxiv.org/abs/1512.00189)

\subsection{S-type Systems}

\subsubsection{TrES-2}

Following \cite{dae09}, TrES-2, also known as Kepler-1, consists of a planet-hosting G0V
primary star with a mass of 1.05 $M_{\odot}$.
The $1.199\pm0.0052 M_{J}$  planet orbits the stellar primary based on a $0.03556\pm0.00075$~au semi-major axis.
The secondary is estimated to have a mass of 0.67 $M_{\odot}$; it is a zero-age main-sequence star. 
The binary separation of the system, which is estimated to be 232$\pm$12~au, ensure that effects by
the secondary star on the primary's habitable environment are largely negligible; therefore, resulting in
conditions akin to a single star.  Thus, S-type HZs should exist around the primary star.
Although the binary eccentricity is unknown, the semi-major-axis of the binary system is larger than
the required value as 23.679~au for S-type GHZ and 19.303~au for S-type RVEM (see Fig.~7).

%(http://www.aanda.org/articles/aa/abs/2009/17/aa10988-08/aa10988-08.html)

\subsubsection{KOI-1257}

Following \cite{san14}, KOI-1257 consists of two main-sequence stars with stellar masses
of 0.99$\pm$0.05 $M_{\odot}$ and 0.70$\pm$0.07 $M_{\odot}$, respectively (see Table~11).
A 1.45$\pm$0.35 $M_{J}$ planet is in an S-type orbit around the primary star with
0.772$\pm$0.045 as eccentricity and 0.382$\pm$0.006 au as semi-major axis. 
The eccentricity of the binary system's orbit is $0.31^{+0.37}_{-0.21}$, and its
semi-major axis is 5.3~au with an uncertainty of 1.3~au.  In this system, the existence of
S-type HZs strongly depends on the binary parameters, which are subject to notable
uncertainties.  At $e_{b}=0.68$, the semi-major axis is required to be larger than 13.919~au
for an S-type GHZ and 11.019~au for RVEM, which means that for this value of $e_b$, there
are no S-type HZs.  However, for the smallest value of $e_{b}$, given as 0.10, the
requirements for the semi-major axis now read 3.899 and 3.085~au for the GHZ and RVEM,
respectively, indicating the existence for both S-type types of HZs (see Fig.~7).
If disregarding the observational uncertainties, the semi-major axis is given as 5.3~au.
This value is larger than 5.386~au for the S-type GHZ to exist, and larger than
4.280~au for the S-type RVEM to exist.

%(https://arxiv.org/abs/1406.6172)

\subsection{Comparison with Other Work}

We also have compared some of our results with previous work that is based
on moderately different methods for the calculations of the HZ limits.  For
S-type and P-type HZs, this work has been given by \cite{hag13} and
\cite{kal13}, respectively.  Figure~8 shows the comparisons for the
systems of Kepler-34, Kepler-35, Kepler-413, and KOI-1257.
Comparisons between results based on GHZ and EVEM climate models are shown
examples, with the observational uncertainties for the eccentricities taken
into account.  The percent differences are calculated as difference between
habitability limits divided by the average.

Regarding Kepler-34, the luminosity of the primary is given as $1.49~L_\odot$
and for the secondary it is $1.28~L_\odot$. The GHZ inner limit has a percent
difference between 3.04\% and 3.14\%, whereas the outer limit's difference
varies between 0.777\% and 0.780\%.  In case of RVEM, the inner limit has
a minimum difference of 0.319\% and a maximum difference of 0.321\%.  The
values for the outer limit are given as 0.390\% and 0.391\%, respectively.
Regarding Kepler-35, the luminosity of the primary is given as $0.94~L_\odot$
and for the secondary it is $0.41~L_\odot$.  The GHZ inner and outer limits
have minimum percent differences of 2.94\% and 0.703\%, and maximum percent
differences of 3.08\% and 0.706\%, respectively.  For the RVEM climate models,
the percent difference for the inner limit and outer limit are close to
0.3\% and 0.5\%, respectively.
Regarding Kepler-413, the luminosity of the primary is given as $0.26~L_\odot$
and for the secondary it is $0.03~L_\odot$.  The inner and outer GHZ limits have
percent differences ranging from 2.82\% to 3.01\% and from 0.562\% and 0.570\%.
RVEM has percent difference between 1.12\% and 1.22\% for the inner limits,
and between 0.518\% and 0.519\% for the outer limits.

KOI-1257, considered as an S-type system, consists of two stars with
luminosities of $1.06~L_\odot$ and $0.14~L_\odot$, respectively.
The minimum and maximum percent differences of the GHZ inner limit
are 3.22\% and 3.73\%, respectively.  The value for the GHZ outer limit
is approximately 0.75\%.  The RVEM inner limit is noted for having
percent differences between 0.053\% and 0.102\%, whereas the RVEM out
limit has a percentage difference of 0.43\% with virtually no variation
regarding the assumed eccentricity of the binary components.
Therefore, in conclusion, our results on obtaining fitting formulae
for the existence of S-type and P-type HZs are unaffected by the choice
of habitability limits as available in the literature.

%%%%%%%%%%%%%%%%%%%%%%%%%%%%%%%%%%%%%%%%%%%%%%%%%%%%%%%%%%%%%%%%%%%%%%%%

\section{Summary and Conclusions}

The aim of this study was the evaluation of the mathematical constraints for
the possibility of HZs in stellar binary systems --- an effort of interest
irrespectively of hitherto planet detections in those systems.  This allowed us
to deduce fitting formulae that permit --- in the framework of the adopted model
--- a straightforward ``yes/no" answer whether HZs exist.  The underlying
mathematical concept is based on the work of Paper~I and II, which 
follows a comprehensive approach for the computation of habitable zones
in binary systems.  The latter includes (1) the consideration of a
joint constraint including orbital stability and a habitable region
for a putative system planet through the stellar radiative energy fluxes
(RHZ) needs to be met; (2) the treatment of different types of HZs as
defined for the Solar System and beyond; (3) the provision of a combined
formalism for the assessment of both S-type and P-type HZs based on
detailed mathematical criteria --- in particular, mathematical criteria
are presented for which kind of system S-type and P-type habitability
is realized; and (4) applications to stellar systems in either circular
or elliptical orbits.  Note that previous less sophisticated fitting
procedures for the existence of HZs in binary systems were given by
\cite{wan16}.

The adopted planetary climate models follow the previous work by
\cite{kop13,kop14}, which allowed us to define the HZs referred to as
GHZ and RVEM, including the definitions of the respective inner and outer
limits.  The inspection of the planetary orbital stability limits follows
the work by \cite{hol99}, which expands on previous results including work
by \cite{dvo86} and \cite{rab88}.  Results on the formation and dynamics
of planets in dual stellar systems compared to single stars have been
given by, e.g., \cite{hag08} and subsequent work.  The fitting formulae
for the existence of the HZs, both regarding  P-type and S-type HZs,
obtained in our study relate the 
 axes $a_{\rm bin}$ between the stellar
binary components and an algebraic expansion for their
orbital eccentricities; they target the limits where the respective HZs
ceases to exist.  For P-type cases, the attained algebraic expansion is
of second order, and for S-type cases, it is of third order, but written
as an exponential exponent.  The various coefficients also depend on
$M_1$ and $M_2$, the masses of the stellar components, which ensures
the comprehensive applicability of the fitting formulae.  The current
version of our methods is aimed at systems of main-sequence stars.

The stellar masses are assumed to range between 0.50 and 1.25~$M_\odot$,
i.e., between spectral type M0V and F6V \citep[e.g.,][]{gra05,man13}.
Thus, the respective stellar luminosities range between 0.036 and 2.15~$L_\odot$.
Therefore, notwithstanding late-type red dwarfs, the kind of stars
for which our approach is applicable comprises more than 95\% of
main-sequence stars \citep[e.g.,][]{kro01,kro02,cha03}.  Furthermore,
detailed tests for our fitting formulae demonstrate that their accuracy
compared to the exact results based on the solutions for the underlying
quartic equations (see Papers~I and II) is better than 5\% in most 
cases, noting that the least accurate results are found for systems
of high eccentricity, i.e., $e_b \gta 0.75$.  In fact, for the vast majority
of cases the accuracy of the fitting formulae is found to be about 1\% or 2\%.

Our method is particularly useful for the quick assessment of observed
systems with relatively large (or poorly known) uncertainties in $a_{\rm bin}$,
$e_b$, $M_1$, and $M_2$, while noting that the latter can play a decisive
role regarding whether or not S-type or P-type HZs exist.  To demonstrate
the applicability of our method, we explored the existence of P-type HZs
for Kepler-34, Kepler-35, Kepler-413, and Kepler-1647 and S-type HZs for
TrES-2 and KOI-1257.  Observational uncertainties of the various system
parameters have been considered as well, which can be relevant for
the outcome.  A good example is KOI-1257, where the existence of the
S-type HZ is strongly affected by the values for both the semi-major axis and
the eccentricity of the stellar motion, which are somewhat uncertain. 
On the other hand, we found that all P-type systems considered
possess P-type HZs irrespectively of the uncertainties in the relevant
observational parameters.  These results are also unaffected by the
planetary climate models.

Generally, the likelihood for the existence of HZs is relatively high
for low values of the $e_b$, but relatively small, or virtually non-existing,
for high values of the $e_b$.  Moreover, the likelihood if a HZs can exist
is also increased if RVEM-type HZs are considered rather than GHZ-type HZs,
as expected.  The fact that high values for $e_b$ decisively reduce the
possibility of S-type habitability has previously been pointed out by, e.g.,
\cite{cun15}.  Our future work will also consider stellar systems
of stars other than main-sequence components and also take into account
future advances about planetary climate models, including the impact of
planetary masses and atmospheric structures.

%%%%%%%%%%%%%%%%%%%%%%%%%%%%%%%%%%%%%%%%%%%%%%%%%%%%%%%%%%%%%%%%%%%%%%%%%%%%%%

\acknowledgments
This work has been supported by the Department of Physics, University
of Texas at Arlington (UTA).  We also appreciate comments by Elke
Pilat-Lohinger about research on orbital stability limits in binary systems. 
Moreover, we wish to thank the anonymous referee for her/his
useful suggestions allowing us to improve the manuscript.

%%%%%%%%%%%%%%%%%%%%%%%%%%%%%%%%%%%%%%%%%%%%%%%%%%%%%%%%%%%%%%%%%%%%%%%%%%%%%%

\clearpage

%\bibliographystyle{natbib}

%%%%%%%%%%%%%%%%%%%%%%%%%%%%%%%%%%%%%%%%%%%%%%%%%%%%%%%%%%%%%%%%%%%%%%%%

\clearpage

%%
%%  FIGURES
%%

%%% *** Fig.1
%%%%%%%%%%%%%%%%%%%%%%%%%%%%%%%%%%%%%%%%%%%%%%%%%%%%%%%%%%%%%%%%%
\begin{figure*} 
\centering
\begin{tabular}{c}
\epsfig{file=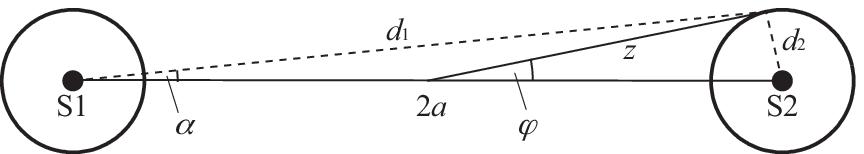,width=0.65\linewidth} \\
\noalign{\bigskip}
\noalign{\bigskip}
\epsfig{file=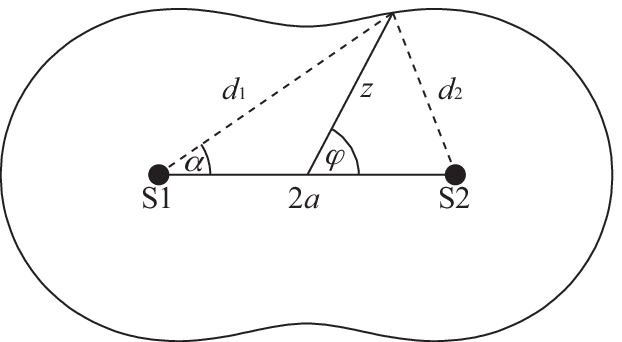,width=0.65\linewidth} \\
\end{tabular}
\caption{
Mathematical set-up of S-type (top) and P-type (bottom) habitable zones
of binary systems as given by the stellar radiative fluxes.  Here $2a$ denotes 
the separation distance between the stellar binary components, corresponding
to the semi-major axis $a_{\rm bin}$ (as used by the observational community)
of the binary system. It is not necessary for the stars S1 and S2
being identical (adopted from Paper~I).
}
\end{figure*}

%+++++++++++++++++++++++++++++++++++++++++++++++++++++++++++++++++++++++

\clearpage

%%% *** Fig.2
%%%%%%%%%%%%%%%%%%%%%%%%%%%%%%%%%%%%%%%%%%%%%%%%%%%%%%%%%%%%%%%%%
\begin{figure*} 
\centering
\begin{tabular}{cc}
\epsfig{file=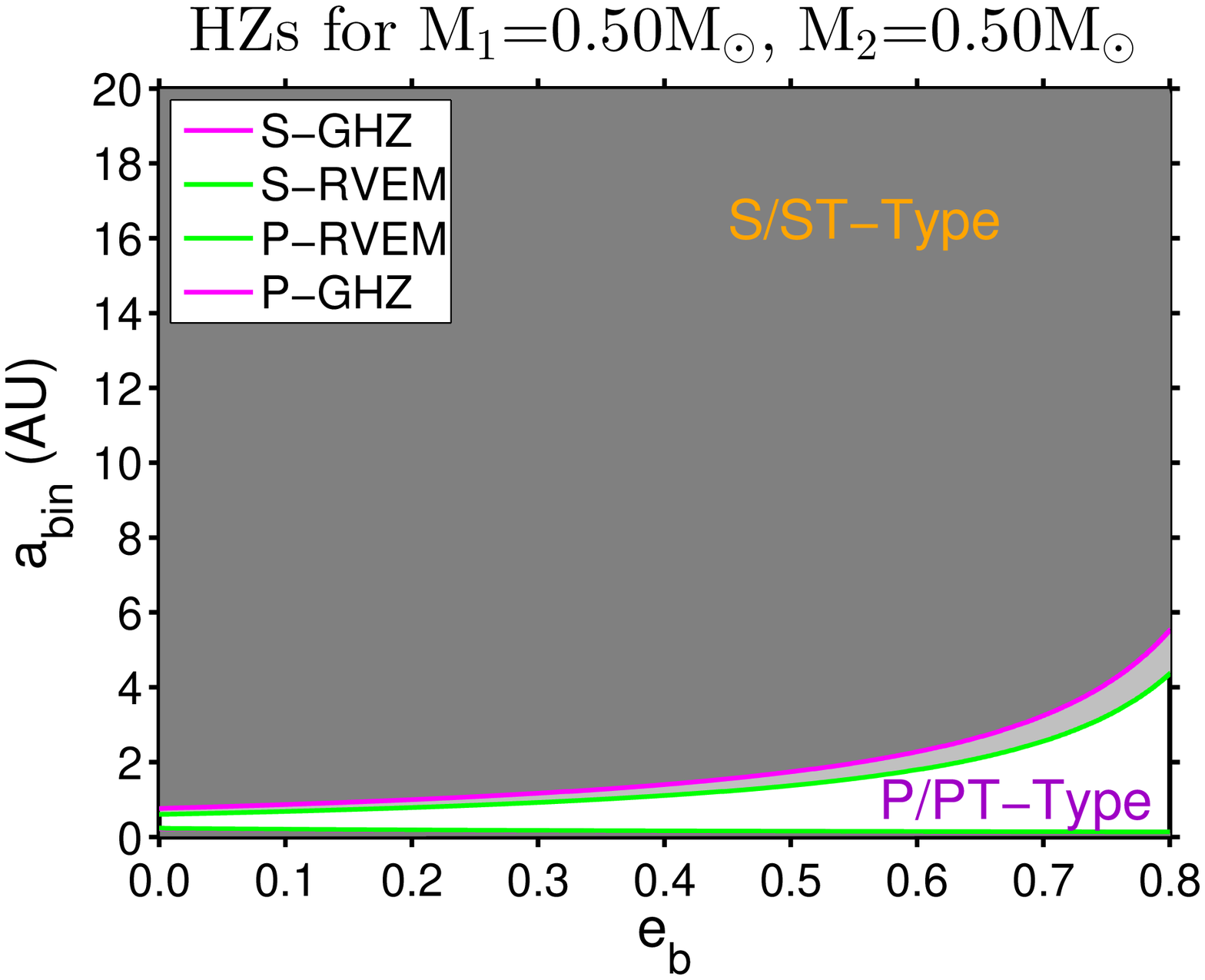,width=0.45\linewidth}
\epsfig{file=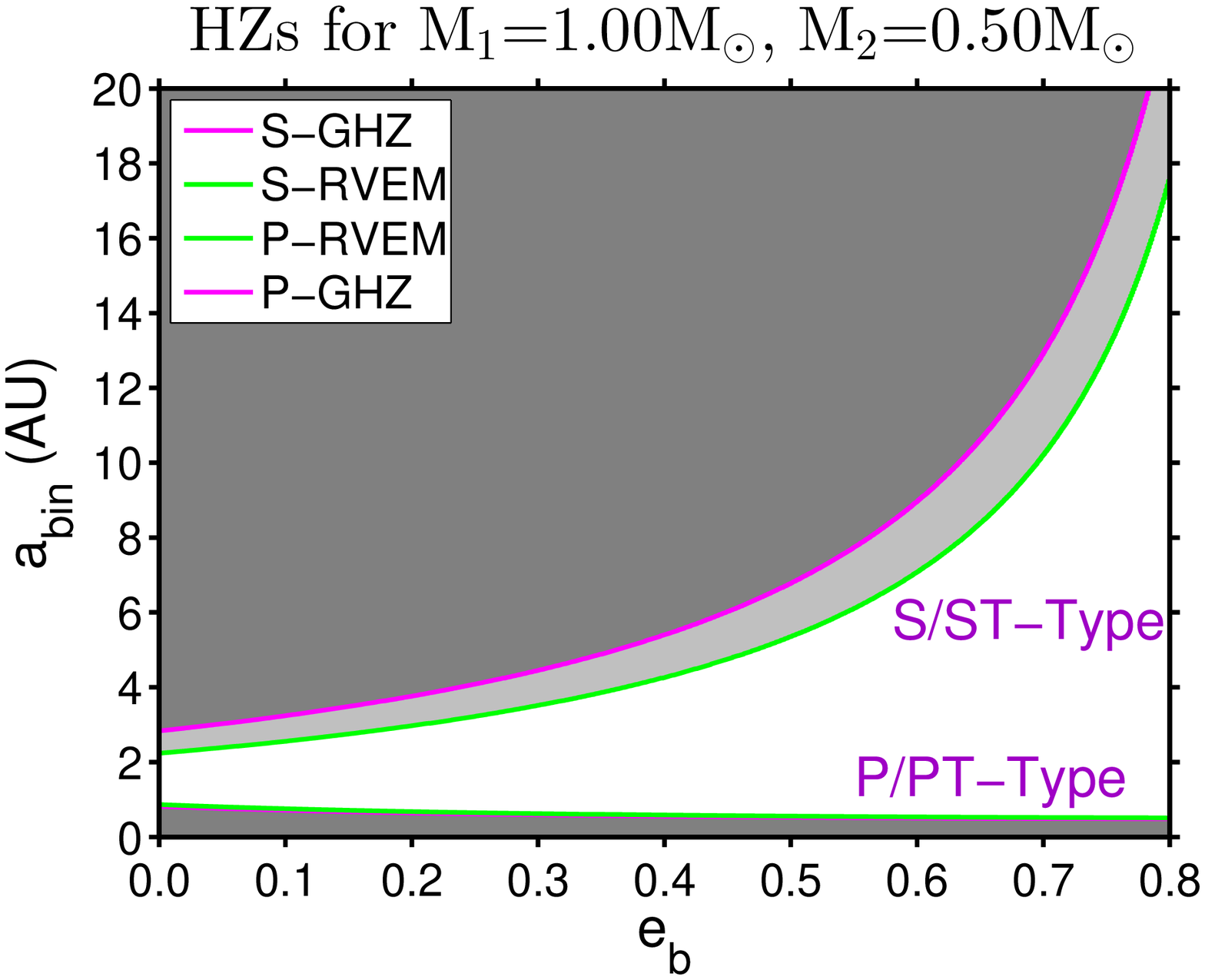,width=0.45\linewidth} \\
\epsfig{file=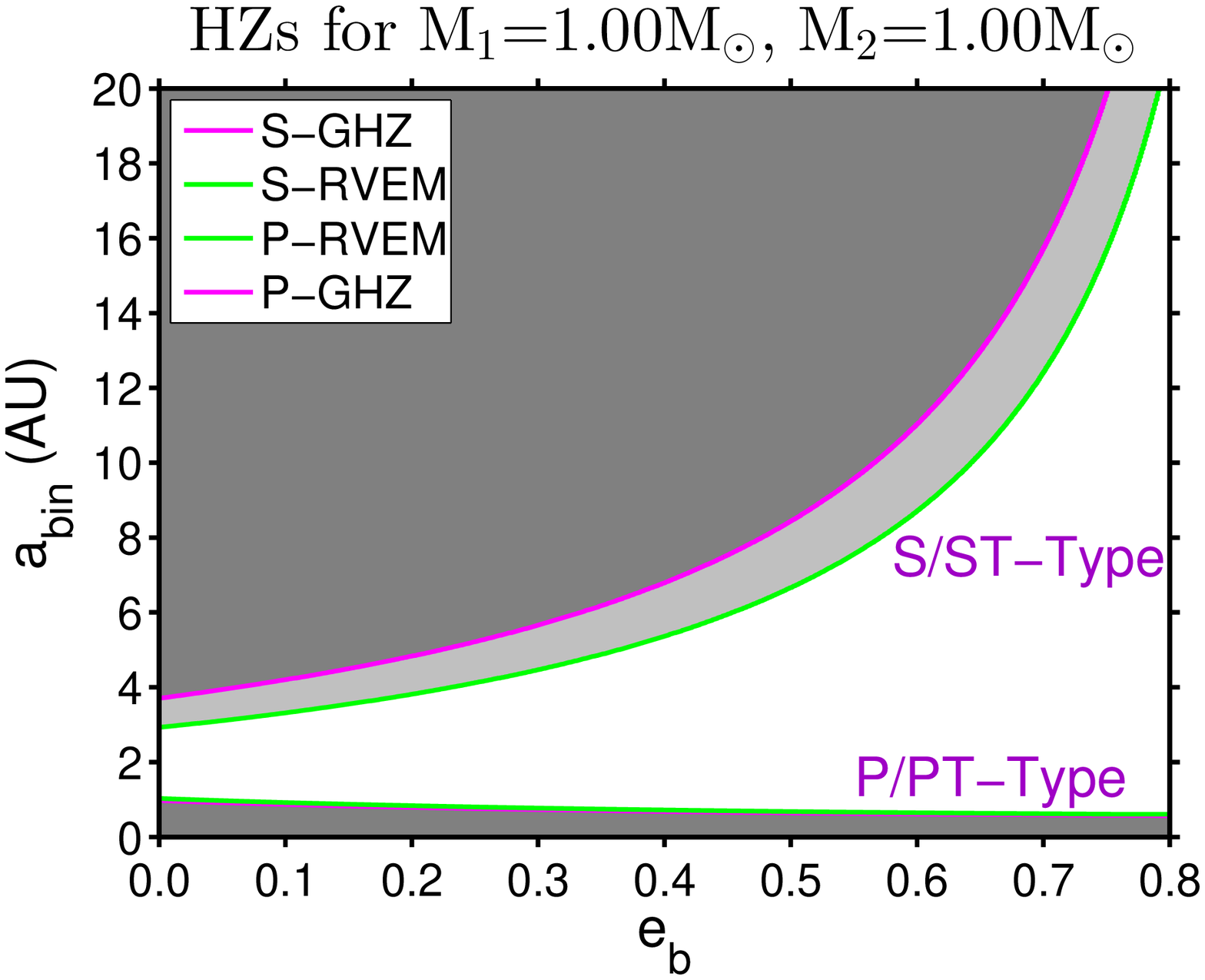,width=0.45\linewidth}
\epsfig{file=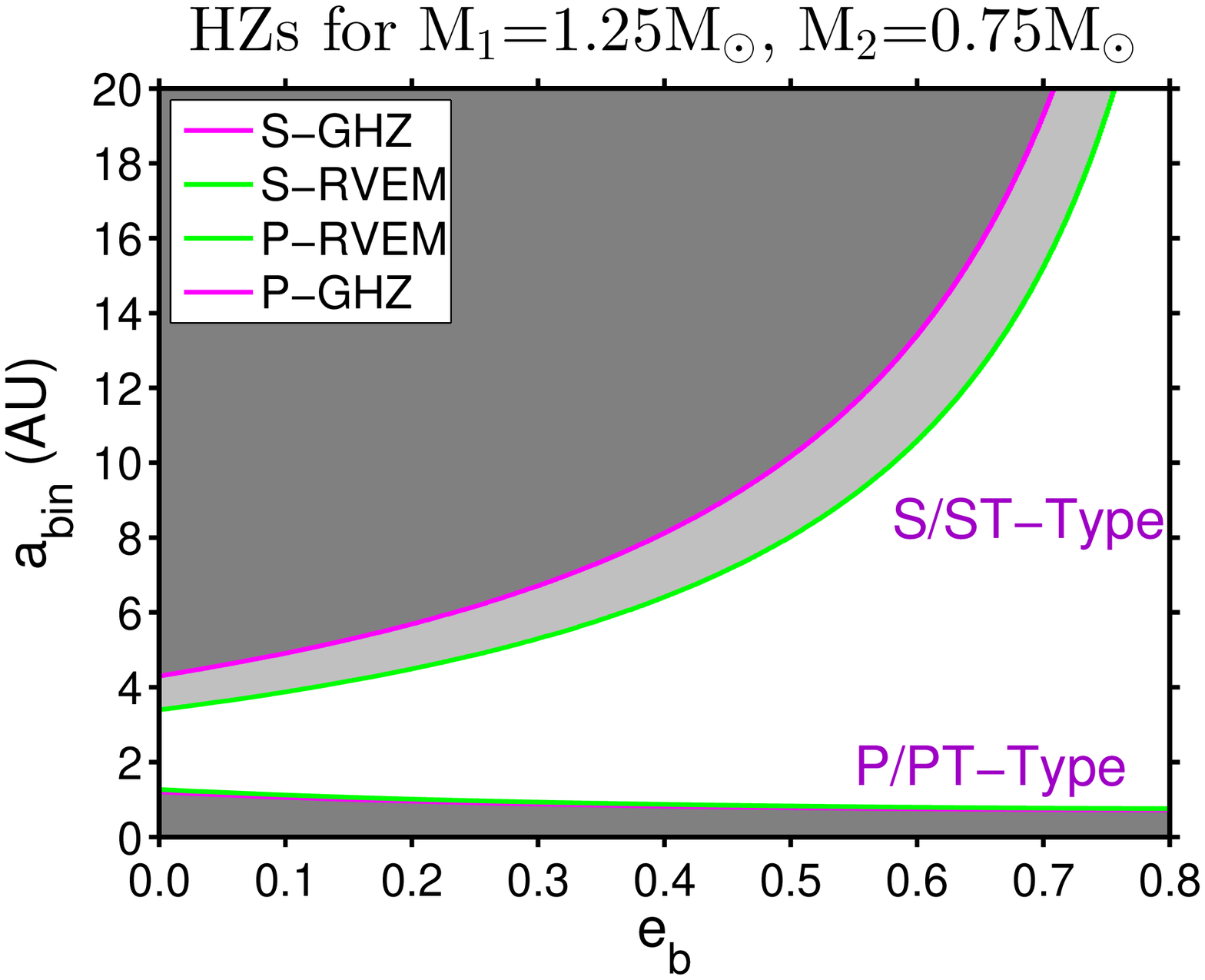,width=0.45\linewidth}
\end{tabular}
\caption{
Required $a_{\rm bin}$ and $e_{b}$ for the GHZ and RVEM to exist regarding selected theoretical
binary systems.  The GHZ can exist when the system parameters are within the gray region.
System parameters fall in either gray or light gray region would allow RVEM to exist.
The magenta and green curves show the critical pairs of values for the GHZ and RVEM
to exist correspondingly.
}
\end{figure*}

%+++++++++++++++++++++++++++++++++++++++++++++++++++++++++++++++++++++++

\clearpage

%%% *** Fig.3
%%%%%%%%%%%%%%%%%%%%%%%%%%%%%%%%%%%%%%%%%%%%%%%%%%%%%%%%%%%%%%%%%
\begin{figure*} 
\centering
\begin{tabular}{cc}
\epsfig{file=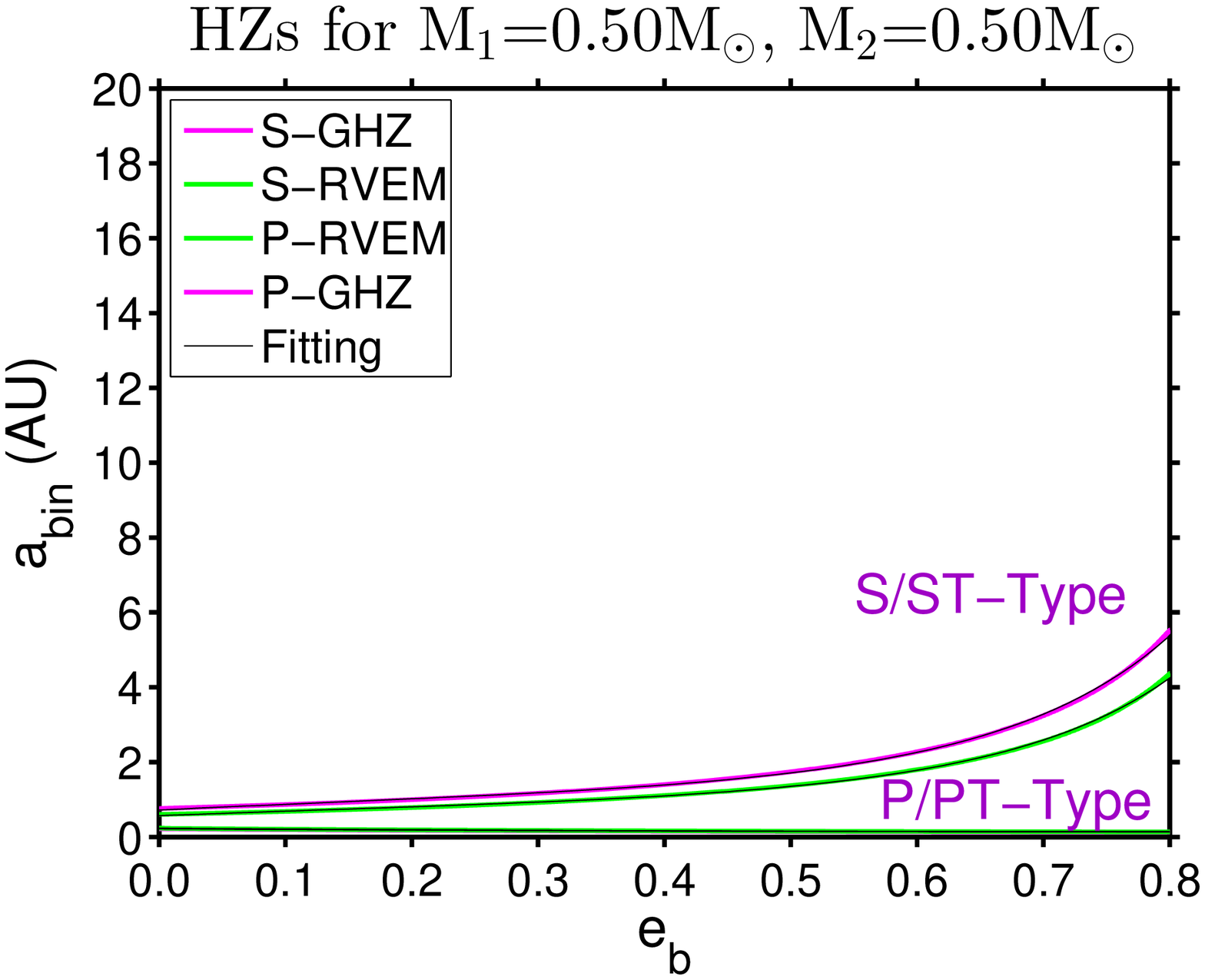,width=0.45\linewidth}
\epsfig{file=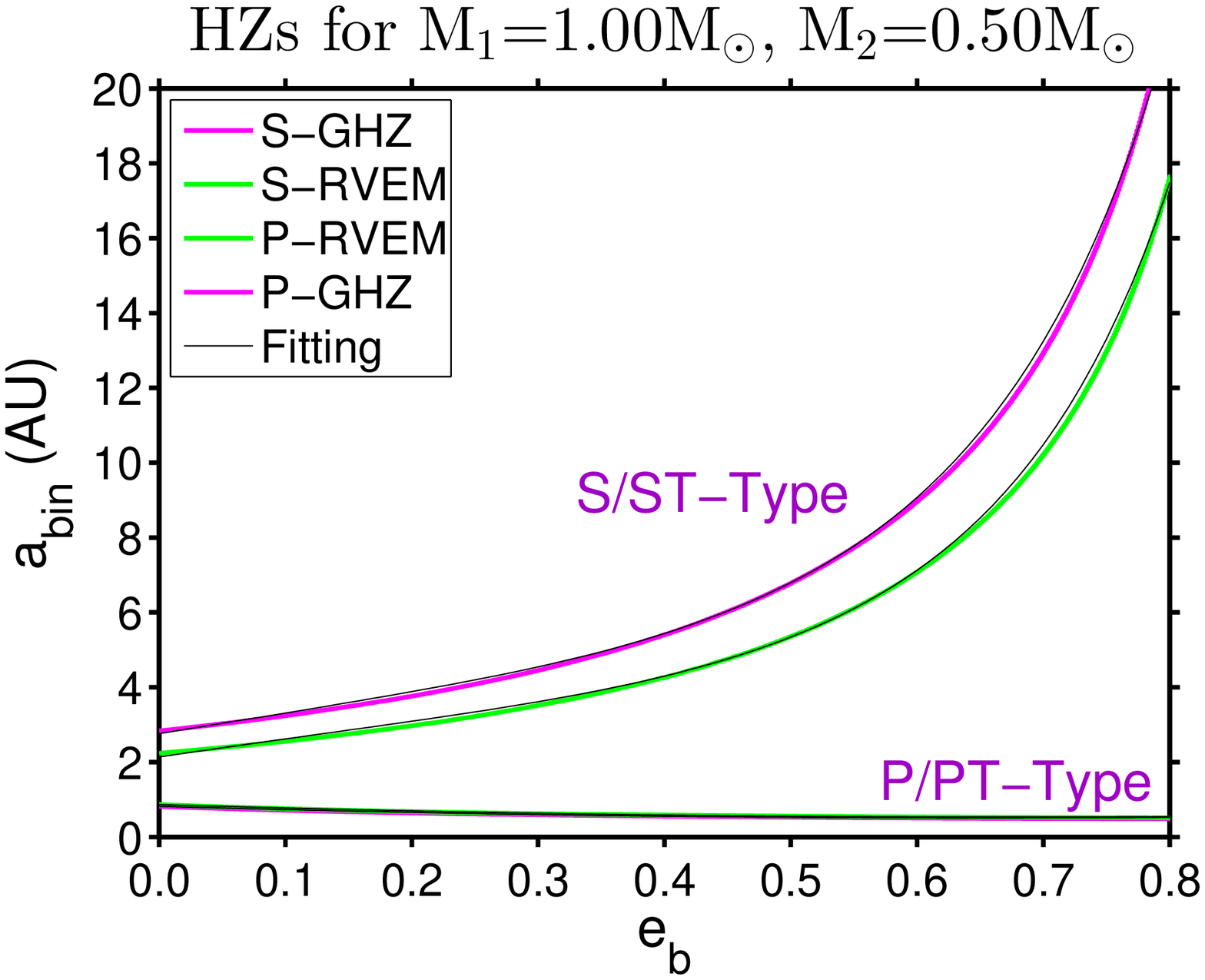,width=0.45\linewidth} \\
\epsfig{file=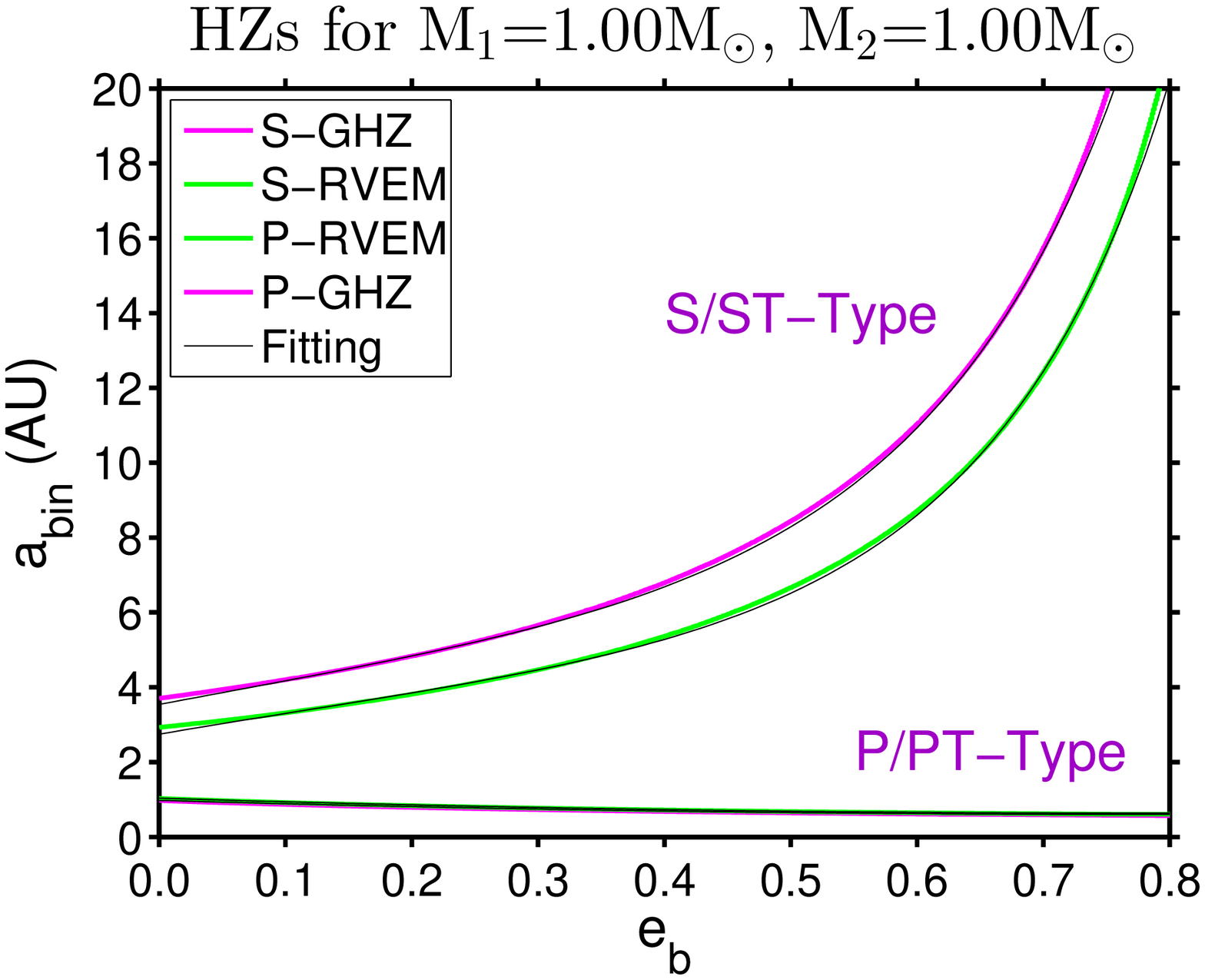,width=0.45\linewidth}
\epsfig{file=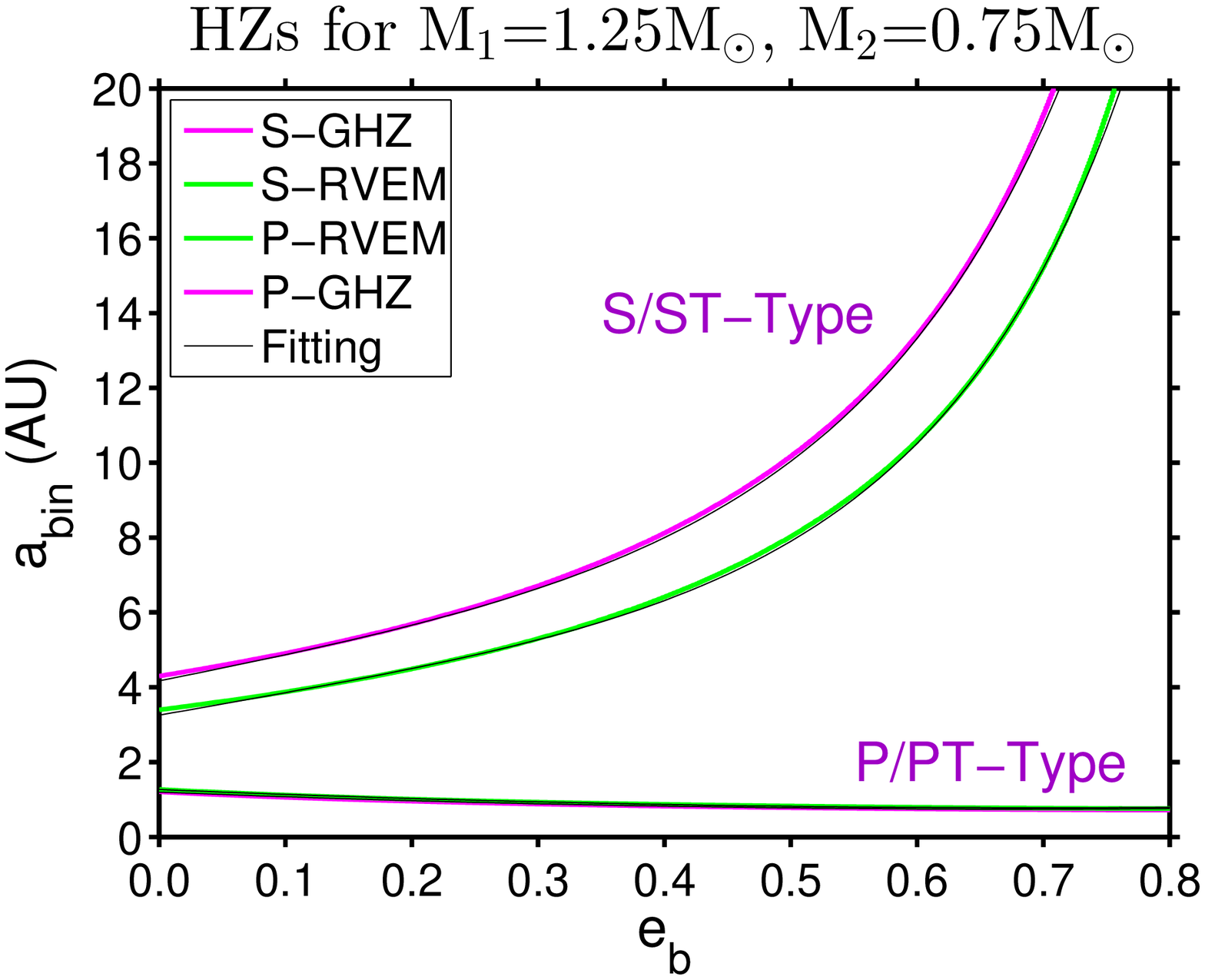,width=0.45\linewidth}
\end{tabular}
\caption{
Fitting of the data for selected theoretical main-sequence stars.
The magenta and green lines represent the boundaries for the GHZ and
the RVEM to exist, respectively.
In each subfigure, the areas beyond the magenta and green curves (top)
identify the existence of the S/ST-type HZs, whereas the areas below the magenta
and green curves (bottom) identify the existence of the P/PT-type HZs.
The thin black curves depict the fitting results for the curve nearby,
and they are virtually indistinguishable from the data curves.
}
\end{figure*}

%+++++++++++++++++++++++++++++++++++++++++++++++++++++++++++++++++++++++

\clearpage

%%% *** Fig.4
%%%%%%%%%%%%%%%%%%%%%%%%%%%%%%%%%%%%%%%%%%%%%%%%%%%%%%%%%%%%%%%%%
\begin{figure*} 
\centering
\begin{tabular}{cc}
\epsfig{file=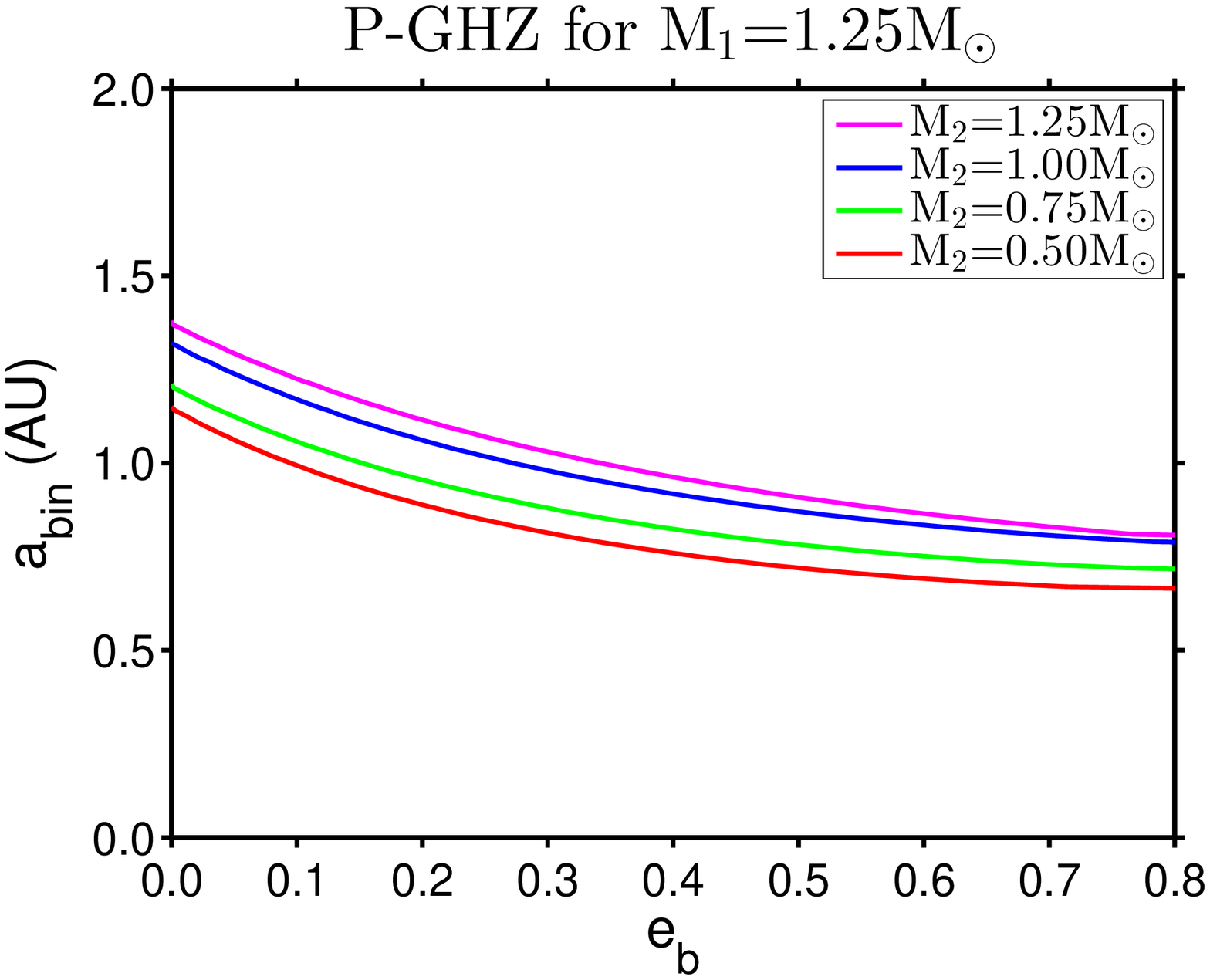,width=0.45\linewidth}
\epsfig{file=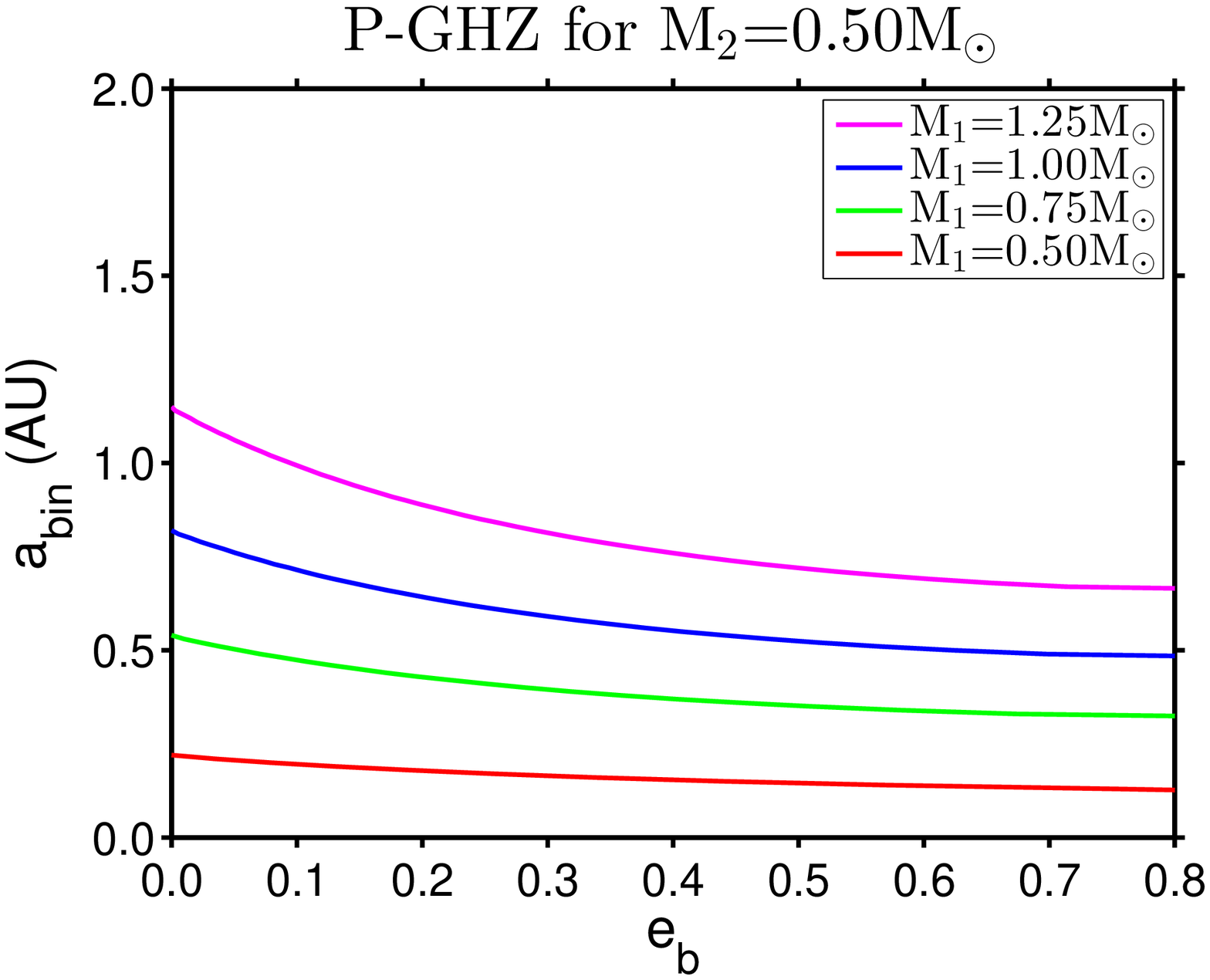,width=0.45\linewidth} \\
\epsfig{file=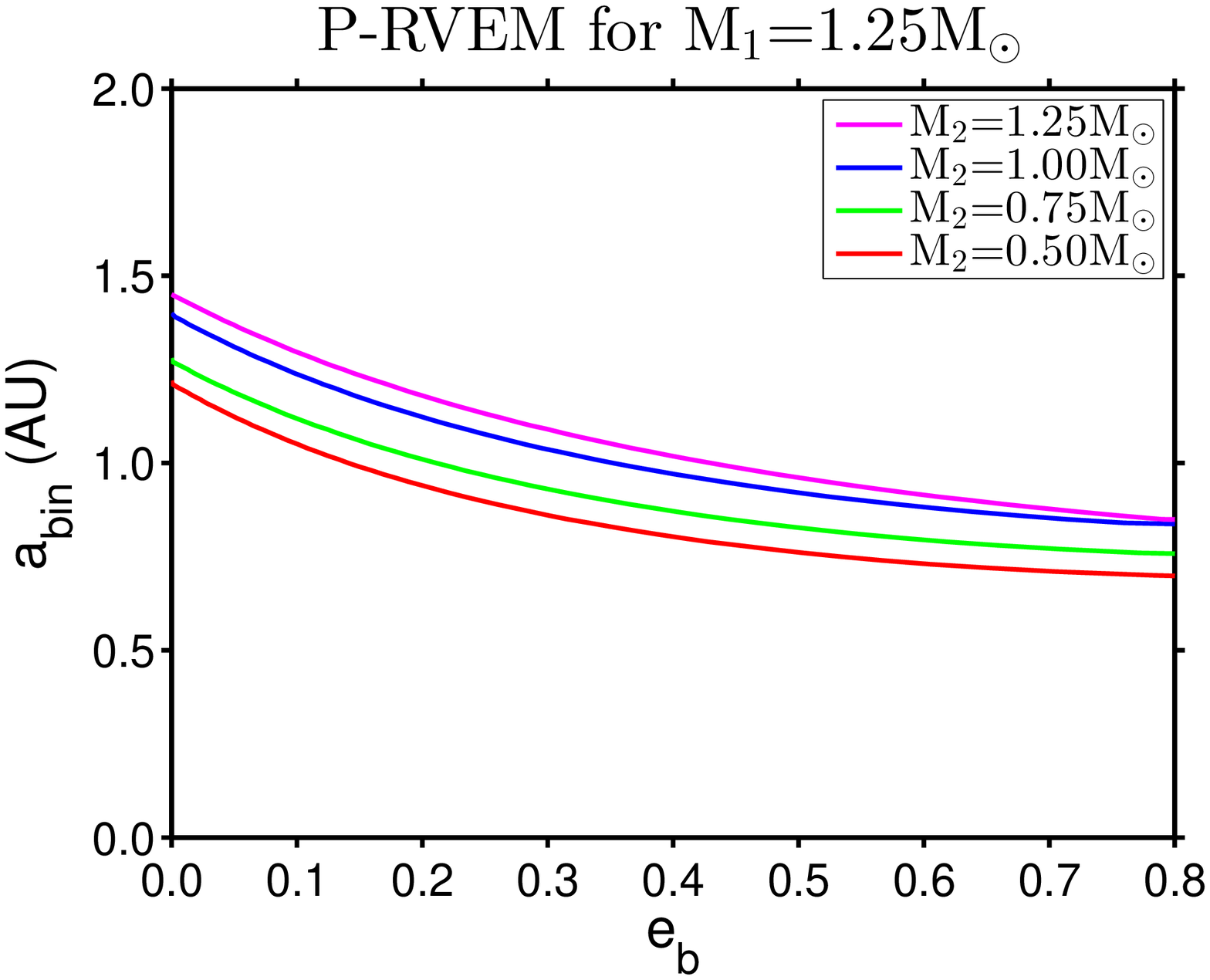,width=0.45\linewidth}
\epsfig{file=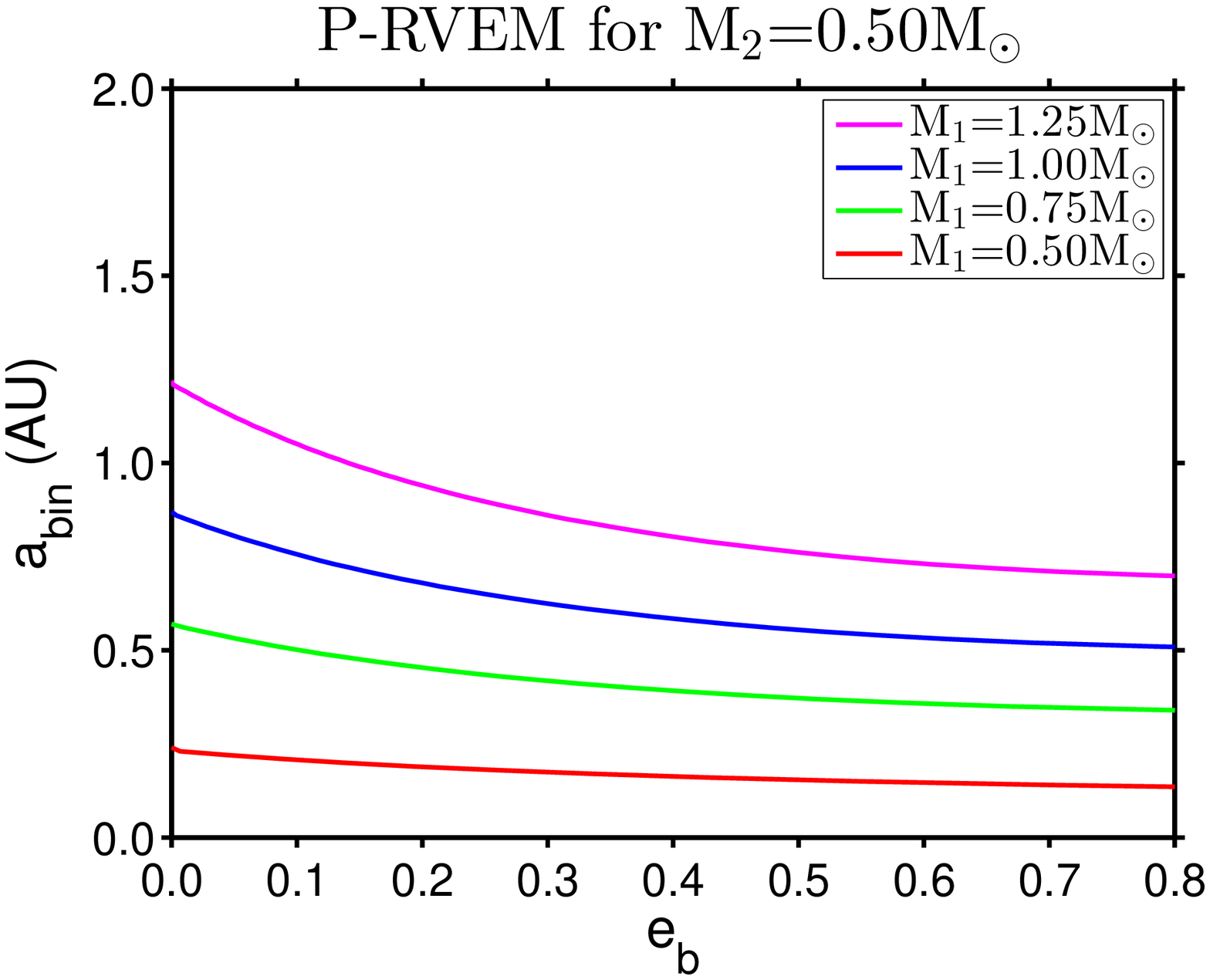,width=0.45\linewidth}
\end{tabular}
\caption{
Results for P-type GHZ and RVEM.  P-type HZs are realized beneath the
respective curve.
}
\end{figure*}

%+++++++++++++++++++++++++++++++++++++++++++++++++++++++++++++++++++++++

\clearpage

%%% *** Fig.5
%%%%%%%%%%%%%%%%%%%%%%%%%%%%%%%%%%%%%%%%%%%%%%%%%%%%%%%%%%%%%%%%%
\begin{figure*} 
\centering
\begin{tabular}{cc}
\epsfig{file=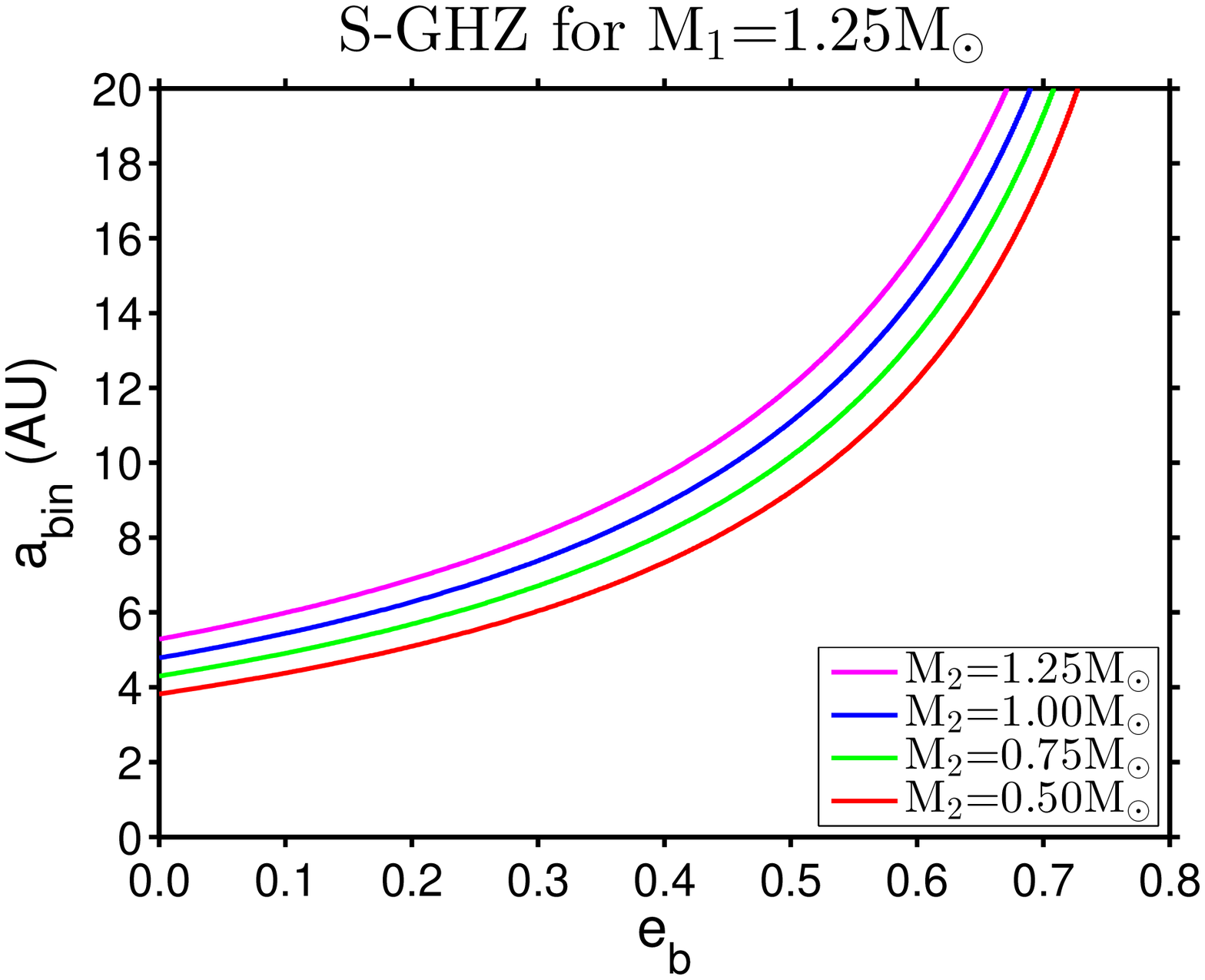,width=0.45\linewidth}
\epsfig{file=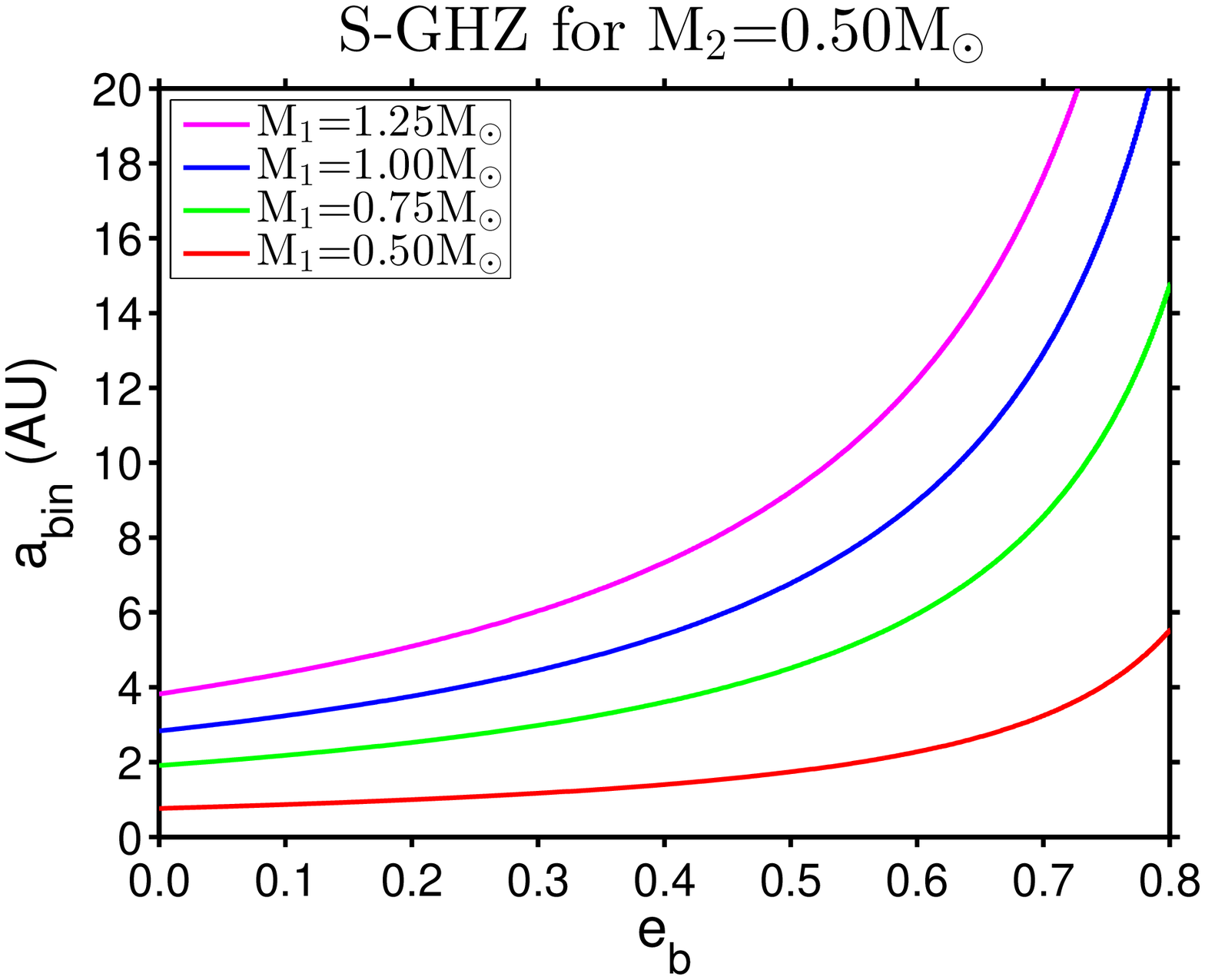,width=0.45\linewidth} \\
\epsfig{file=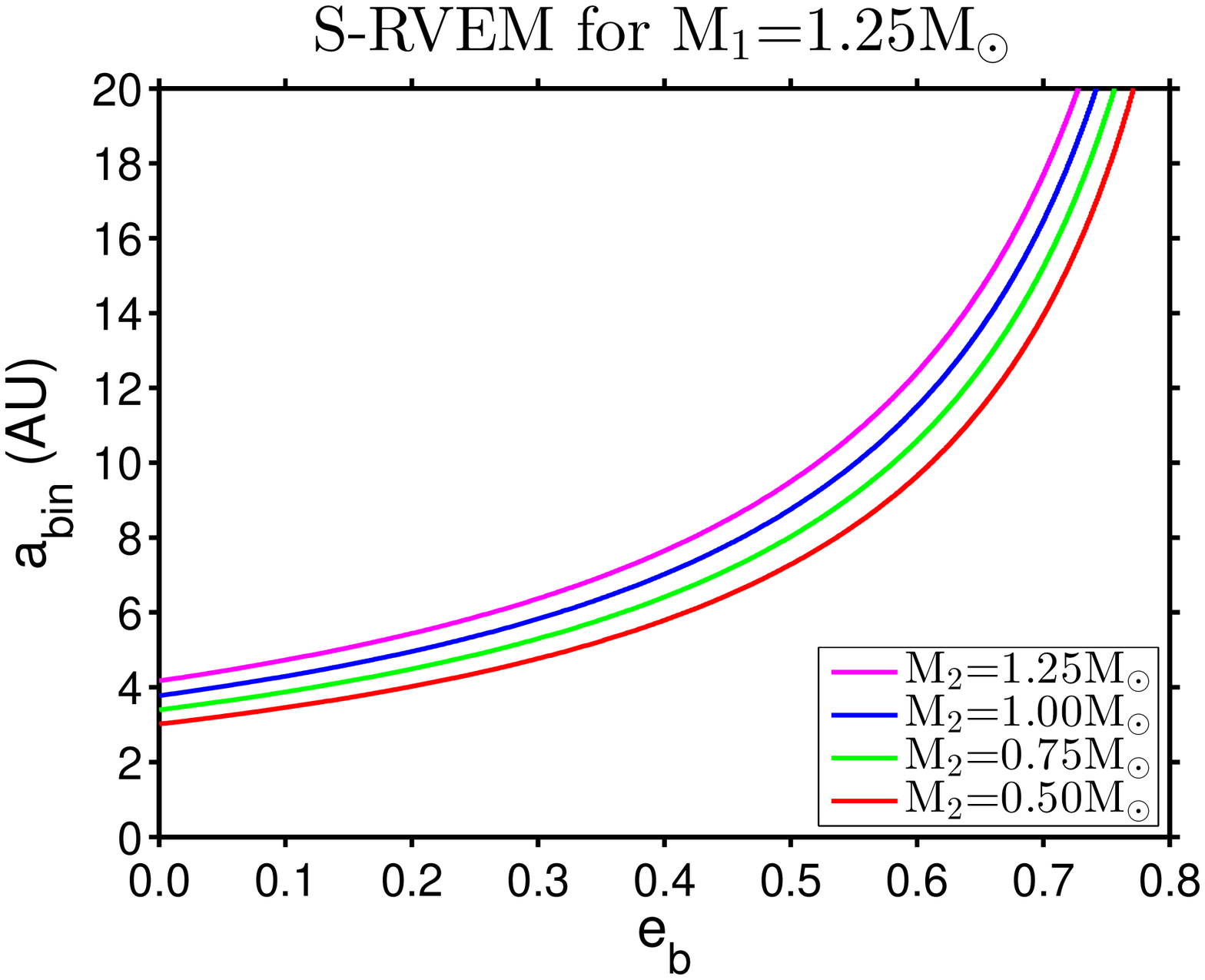,width=0.45\linewidth}
\epsfig{file=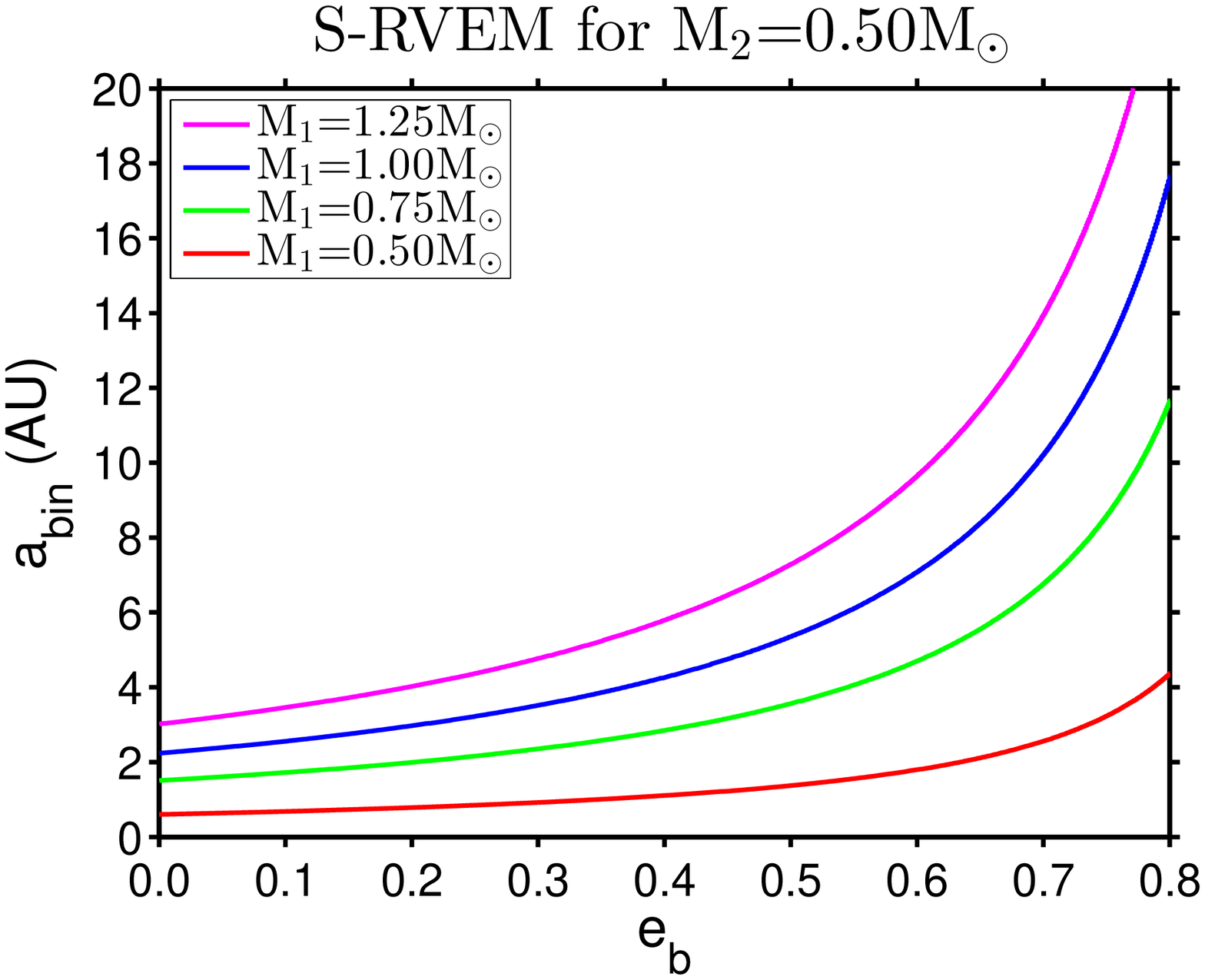,width=0.45\linewidth}
\end{tabular}
\caption{
Results for S-type GHZ and RVEM.  S-type HZs are realized
above the respective curve.
}
\end{figure*}

%+++++++++++++++++++++++++++++++++++++++++++++++++++++++++++++++++++++++

\clearpage

%%% *** Fig.6
%%%%%%%%%%%%%%%%%%%%%%%%%%%%%%%%%%%%%%%%%%%%%%%%%%%%%%%%%%%%%%%%%
\begin{figure*} 
\centering
\begin{tabular}{cc}
\epsfig{file=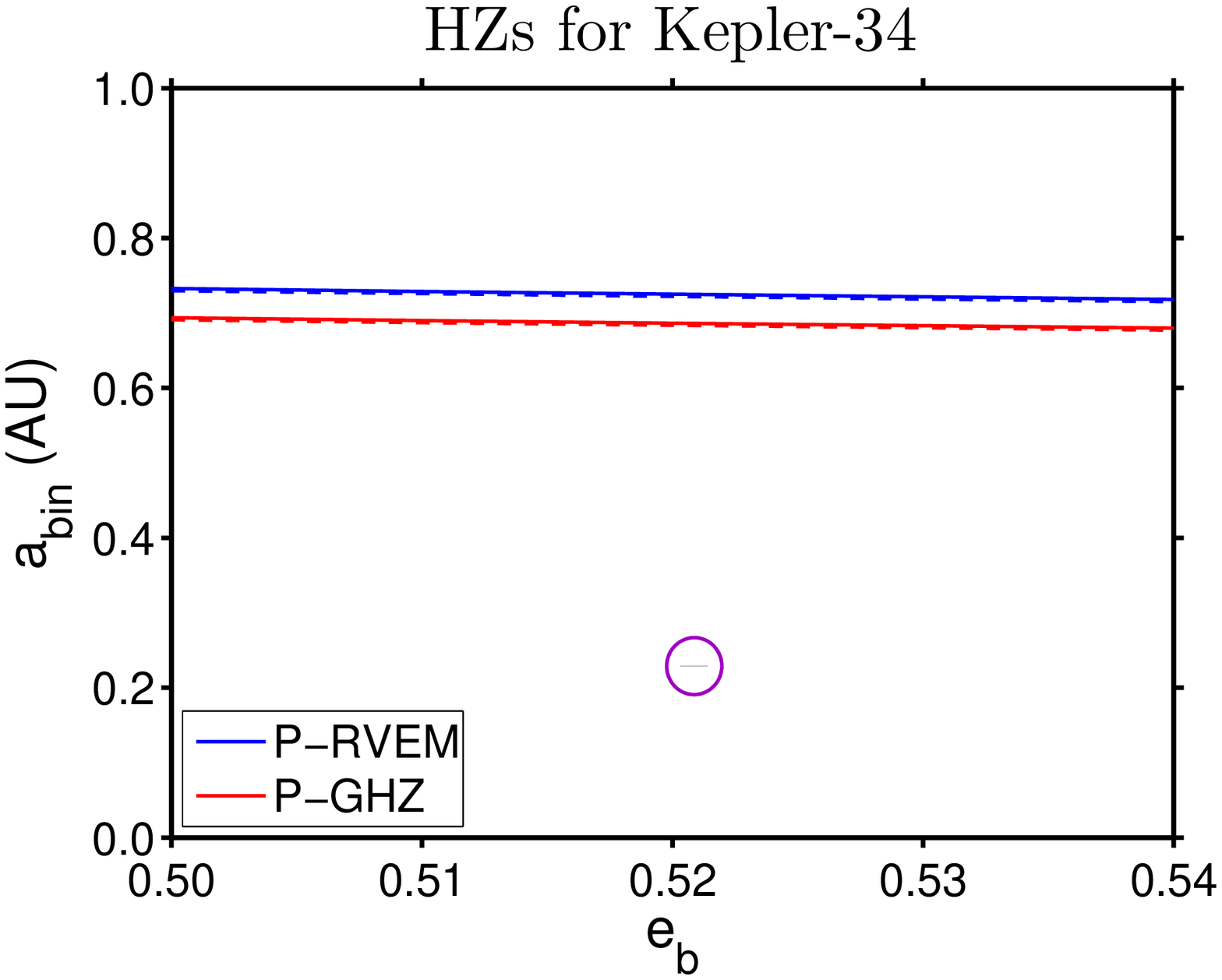,width=0.45\linewidth}
\epsfig{file=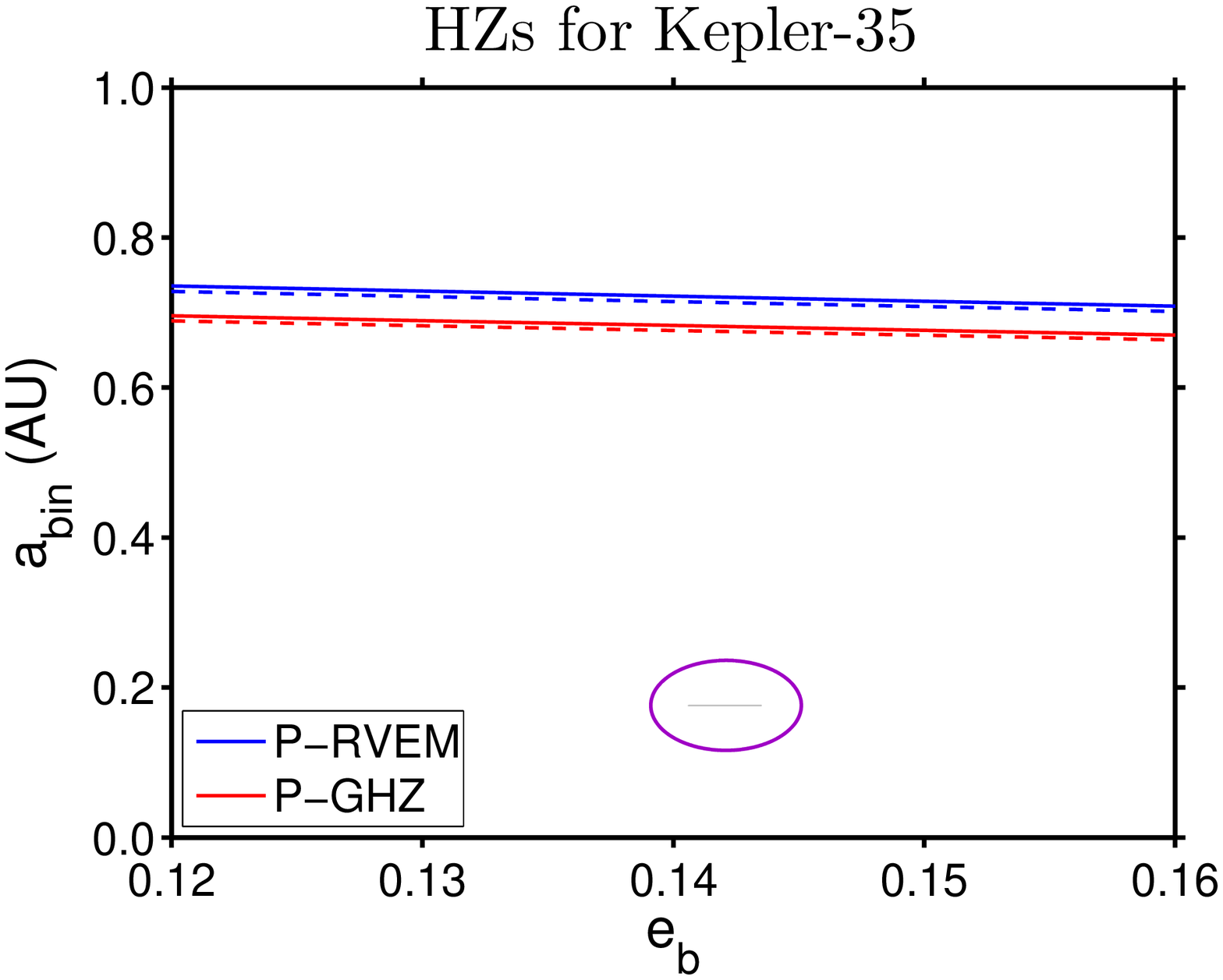,width=0.45\linewidth} \\
\epsfig{file=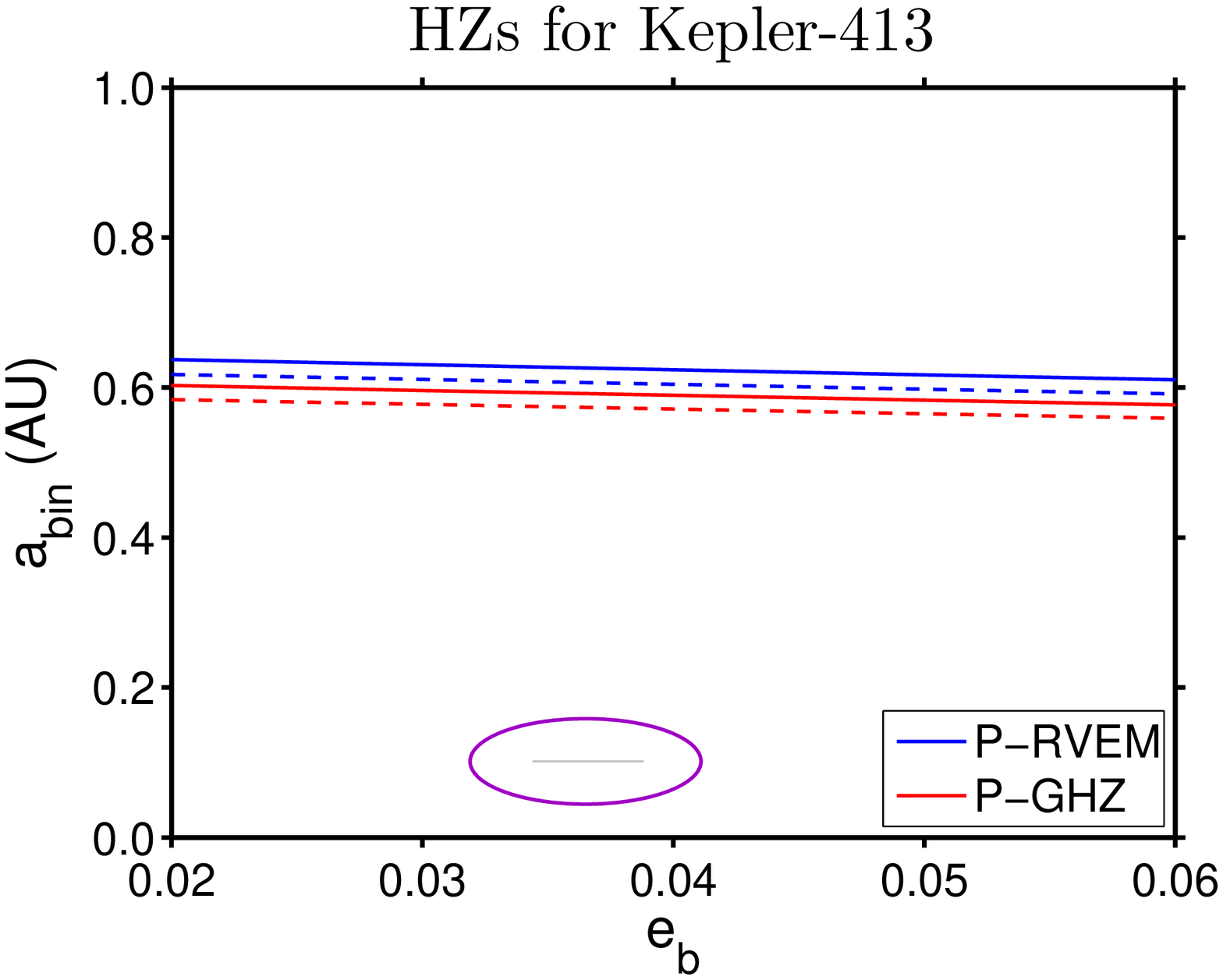,width=0.45\linewidth}
\epsfig{file=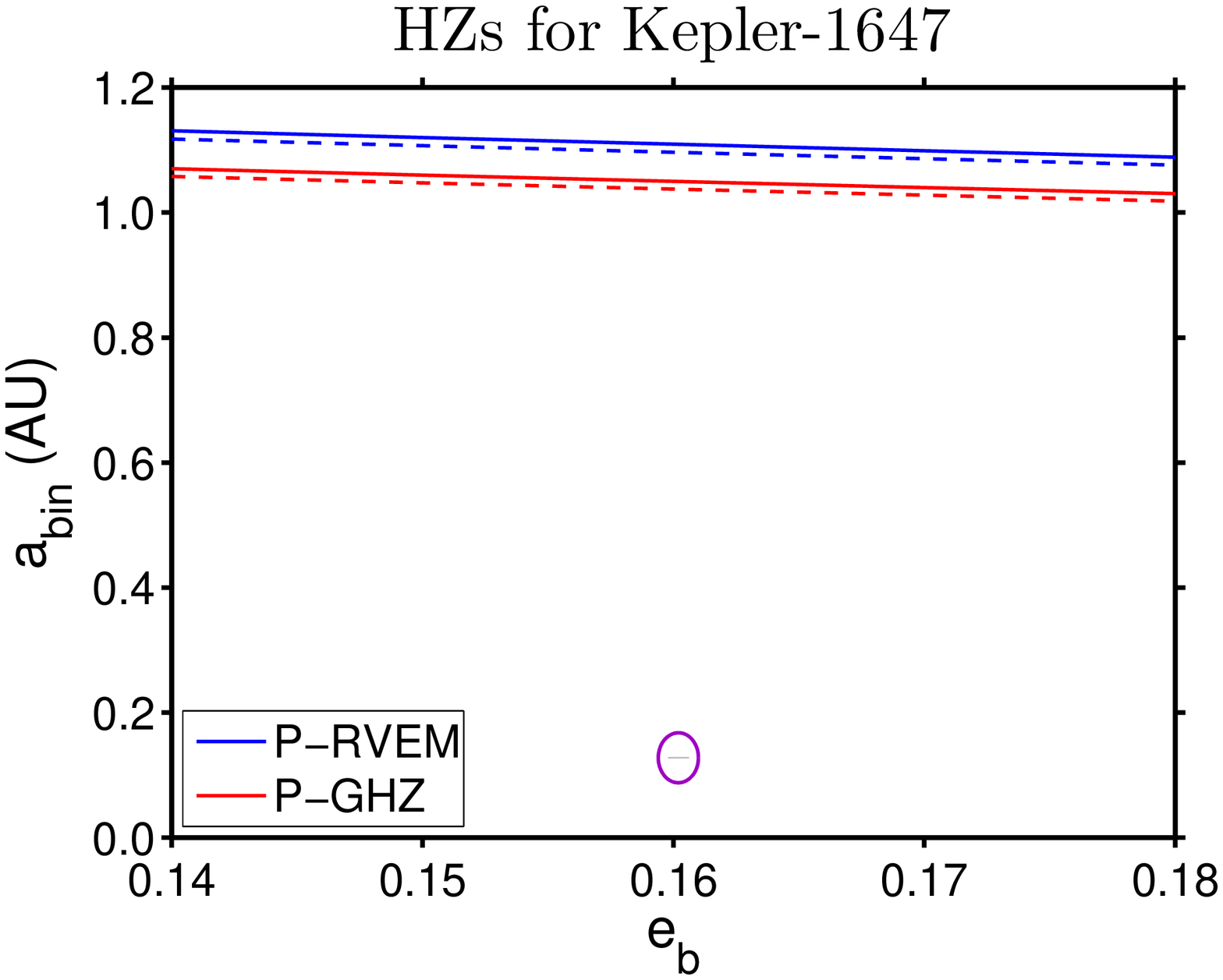,width=0.45\linewidth}
\end{tabular}
\caption{
The red and blue curves are the fitting results that show the maximum $a_{\rm bin}$ for
P-type HZs to exist; i.e., P-type HZs are possible below these curves.
The dashed lines are the results considering the uncertainties in the stellar masses.
The gray domains indicate the indicated $a_{\rm bin}$ and $e_b$ values for the respective
stellar systems with the observational uncertainties taken into account
(the purple ellipses are placed to enhance the domains' visibility).
}
\end{figure*}

%+++++++++++++++++++++++++++++++++++++++++++++++++++++++++++++++++++++++

\clearpage

%%% *** Fig.7
%%%%%%%%%%%%%%%%%%%%%%%%%%%%%%%%%%%%%%%%%%%%%%%%%%%%%%%%%%%%%%%%%
\begin{figure*} 
\centering
\begin{tabular}{cc}
\epsfig{file=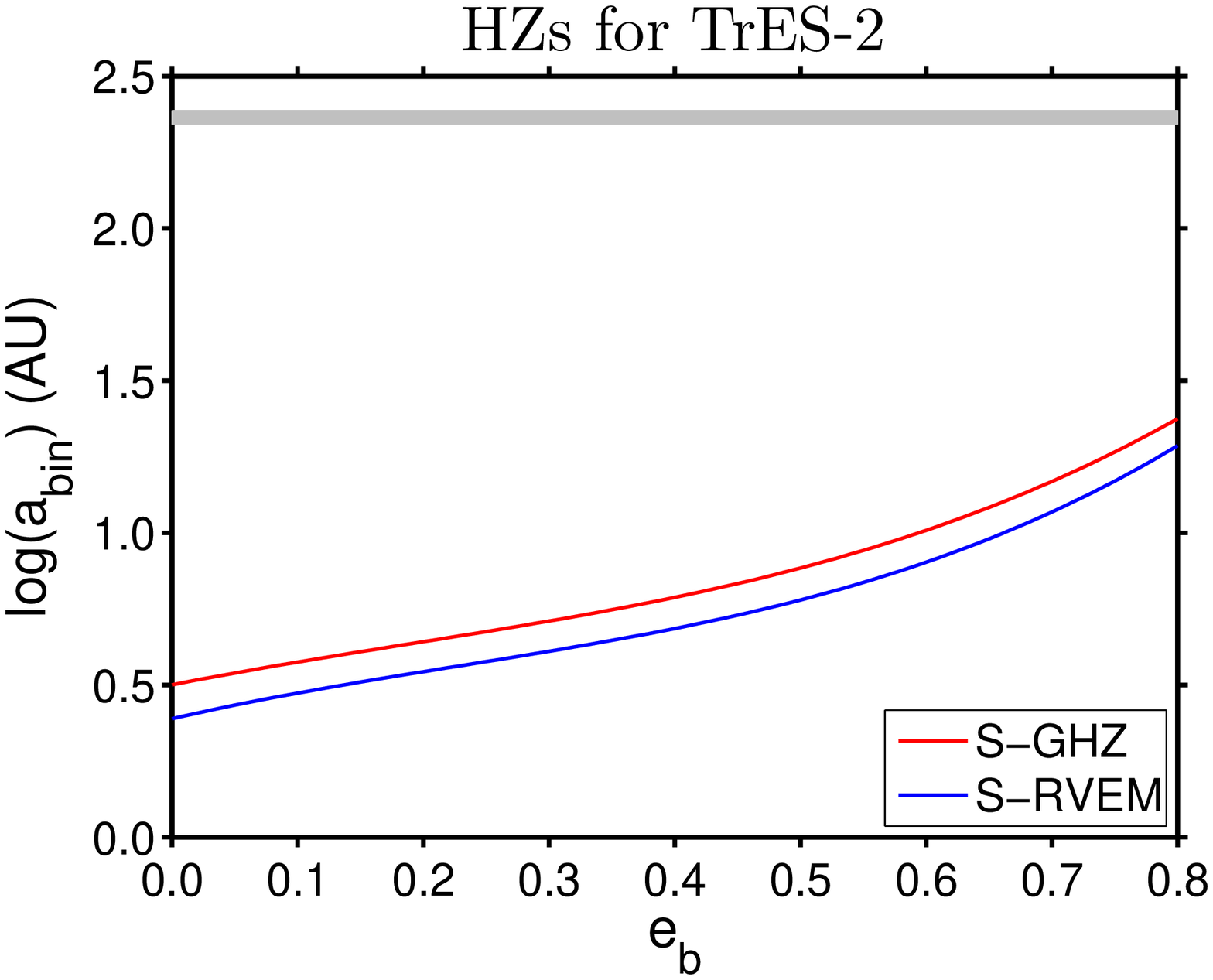,width=0.45\linewidth}
\epsfig{file=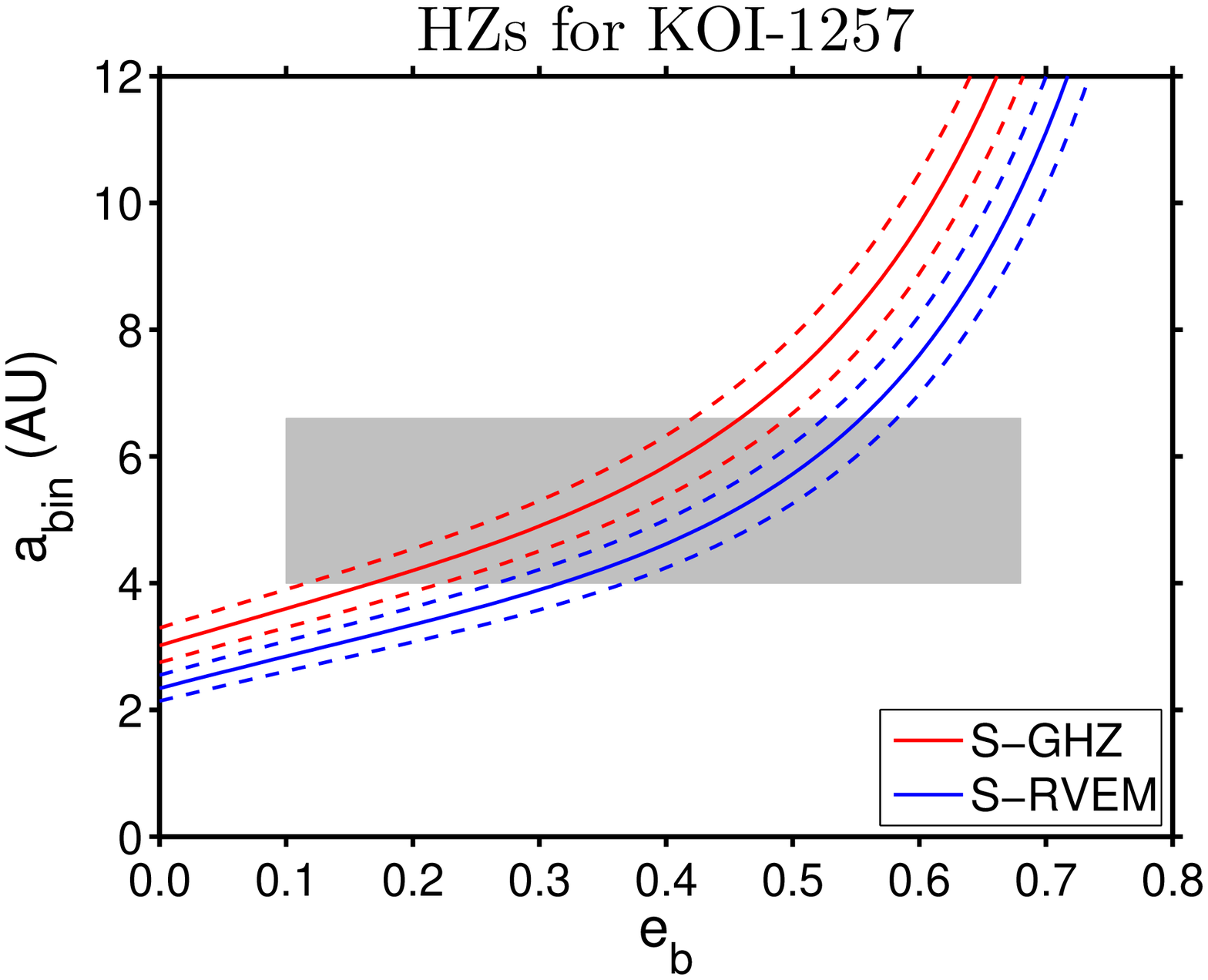,width=0.45\linewidth}
\end{tabular}
\caption{
The red and blue curves are the fitting results that show the minimum $a_{\rm bin}$ for
S-type HZs to exist; i.e., S-type HZs are possible above these curves.
The dashed lines are the results considering the uncertainties in the stellar masses.
The gray domains indicate the indicated $a_{\rm bin}$ and $e_b$ values for the respective
stellar systems with the observational uncertainties taken into account.
A logarithmic scale $y$-scale is used for TrES-2 because of the system's
very large semi-major axis.
}
\end{figure*}

%+++++++++++++++++++++++++++++++++++++++++++++++++++++++++++++++++++++++

\clearpage

%%% *** Fig.8
%%%%%%%%%%%%%%%%%%%%%%%%%%%%%%%%%%%%%%%%%%%%%%%%%%%%%%%%%%%%%%%%%
\begin{figure*} 
\centering
\begin{tabular}{cc}
\epsfig{file=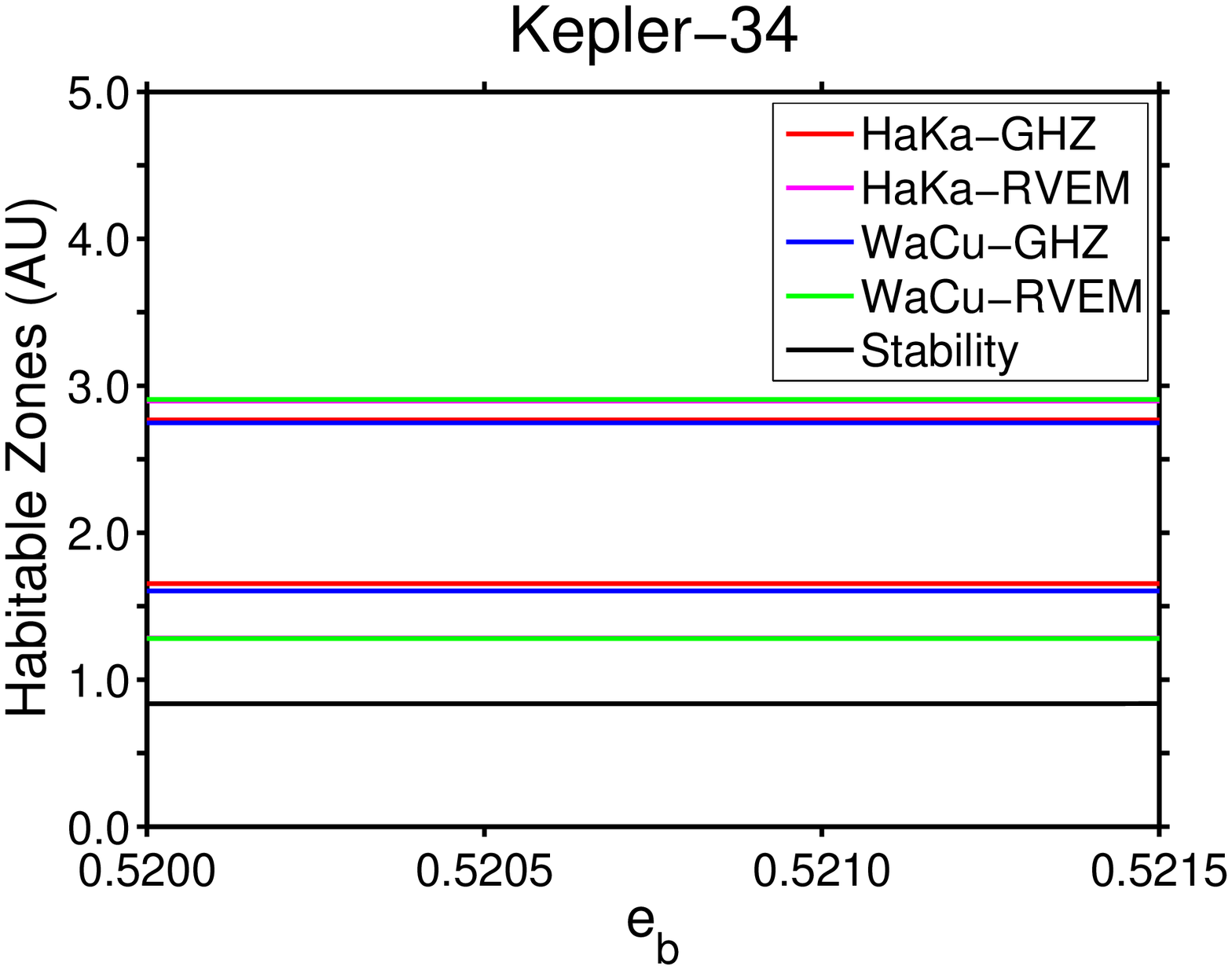, width=0.45\linewidth}
\epsfig{file=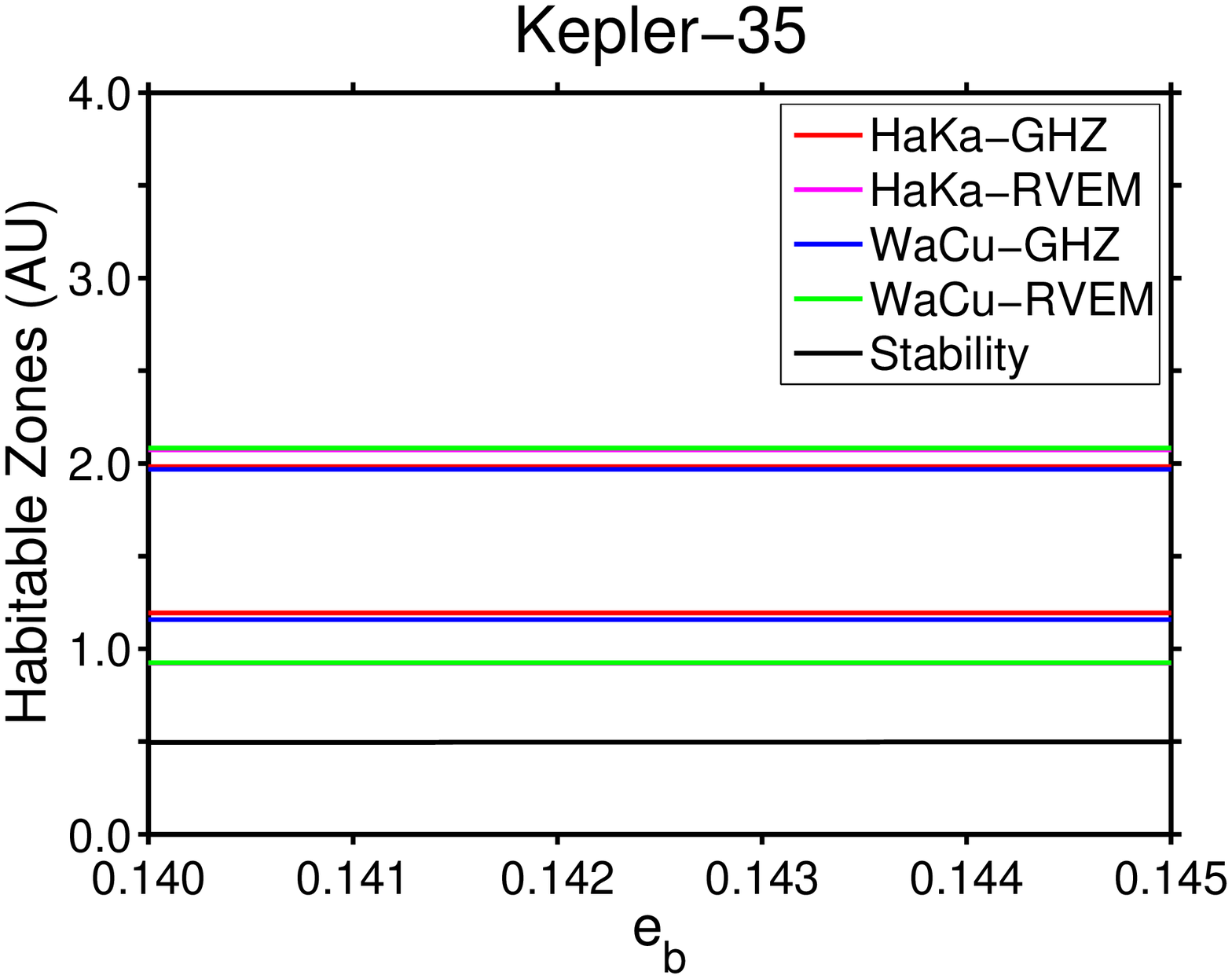, width=0.45\linewidth} \\
\epsfig{file=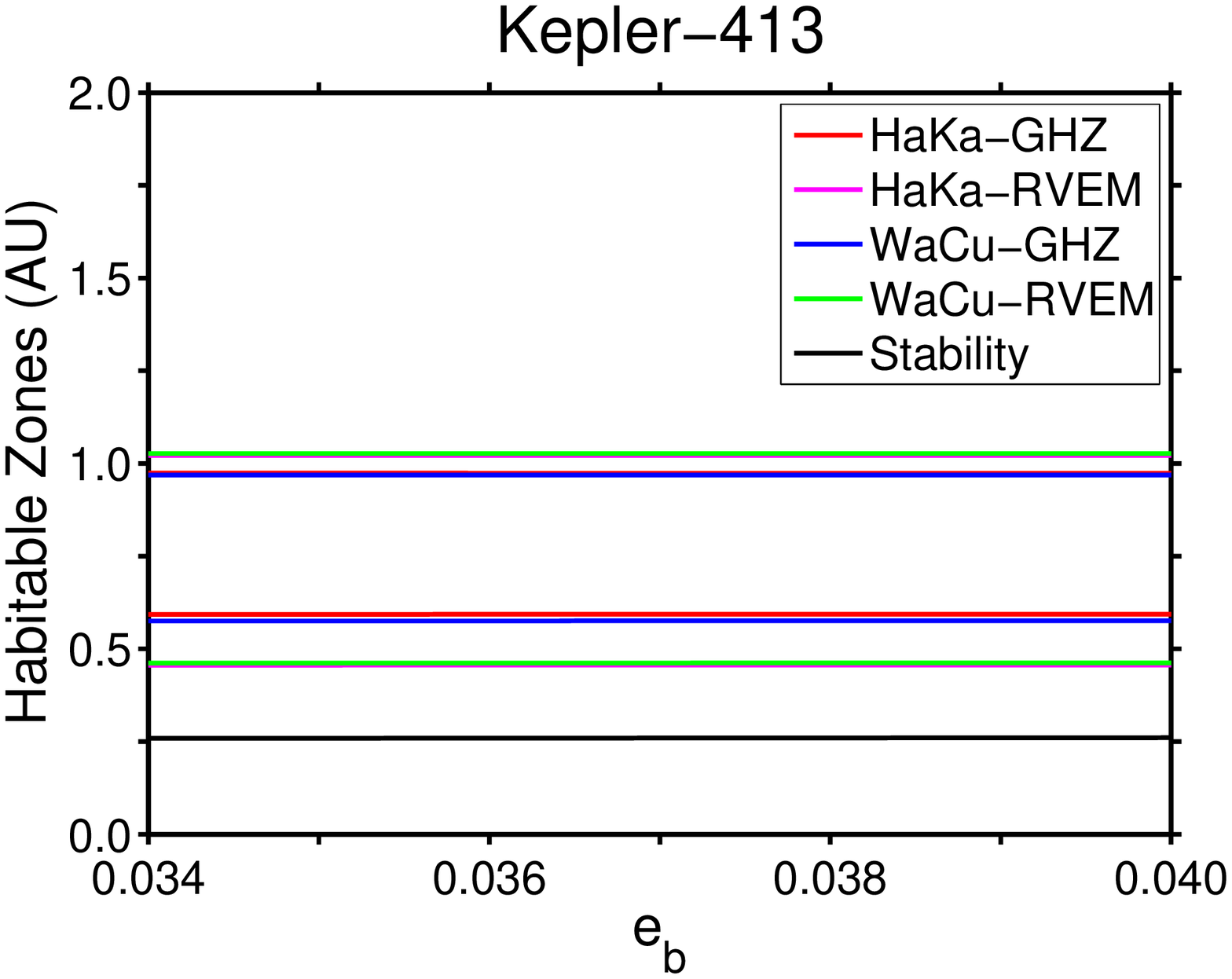,width=0.45\linewidth}
\epsfig{file=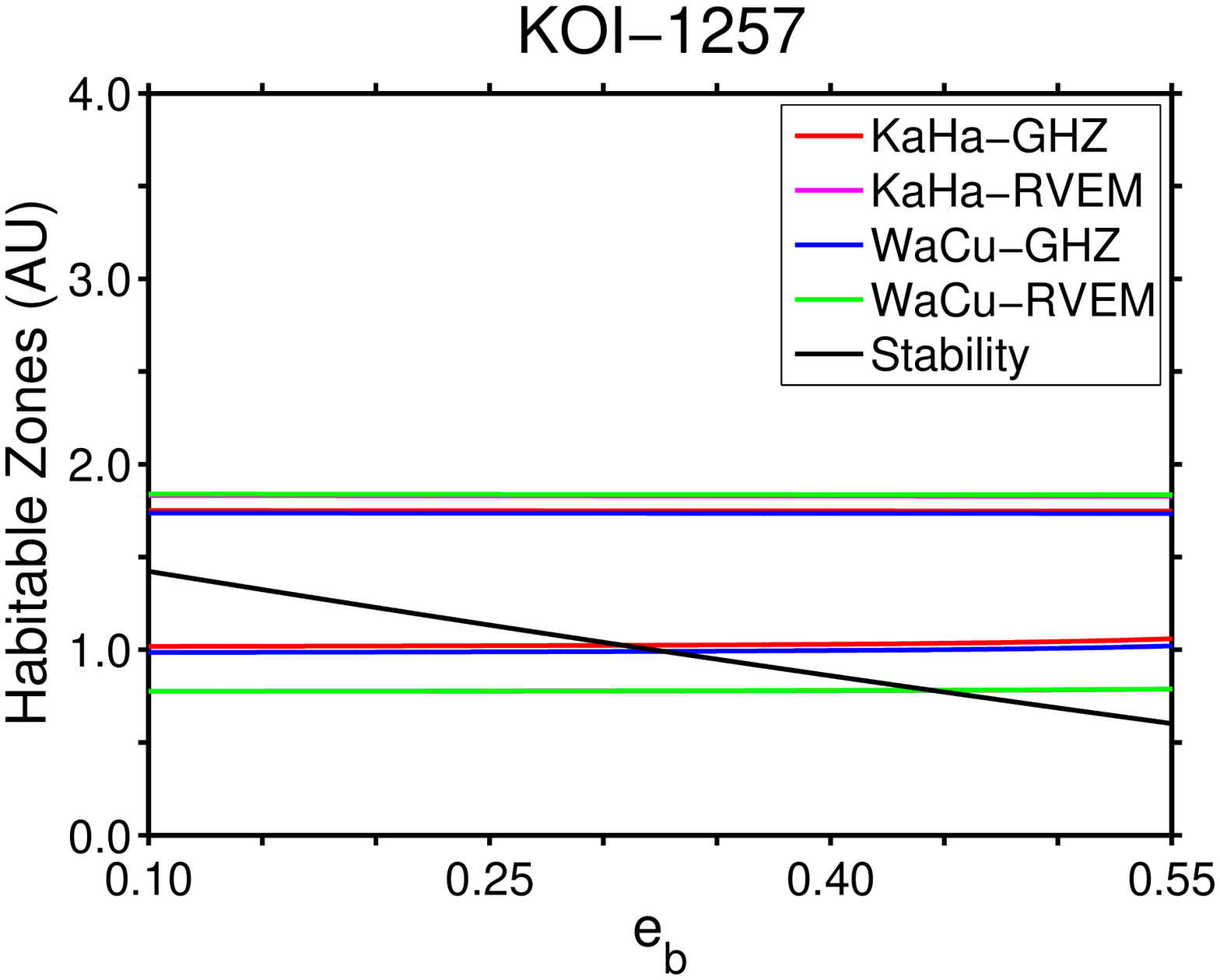,  width=0.45\linewidth}
\end{tabular}
\caption{
Comparisons of our results (WaCu) for Kepler-34, Kepler-35,
Kepler-413, and KOI-1257 regarding the existence of HZs
with previous work pertaining to the calculations of HZ limits.
That work has been given by \cite{hag13} for S-type HZs and by
\cite{kal13} for P-type HZs, thus denoted as HaKa and KaHa,
respectively.  Results are given for GHZ and RVEM climate models.
}
\end{figure*}

%+++++++++++++++++++++++++++++++++++++++++++++++++++++++++++++++++++++++

\clearpage

%%
%%  TABLES
%%

%% *** Table 1
%%%%%%%%%%%%%%%%%%%%%%%%%%%%%%%%%%%%%%%%%%%%%%%%%%%%%%%%%%%%%%%%%%%%%%%%
\begin{deluxetable}{lccccl}
\tablecaption{Habitability Limits for the Solar System}
\tablewidth{0pt}
\tablehead{
Description & Indices & \multicolumn{3}{c}{Models} & This work     \\
       ...  & $l$     & \multicolumn{2}{c}{Kas93}  & Kop1314 & ... \\
\noalign{\smallskip}
\hline
\noalign{\smallskip}
		...  & ... & 5700~K & 5780~K & 5780~K & ... \\
		...  & ... & (au)   & (au)   & (au)   & ... 
}
\startdata
		Recent Venus			&  1  &  0.75    &  0.77		& 0.750	& RVEM Inner Limit \\
		Runaway greenhouse effect	&  2  &  0.84    &  0.86		& 0.950	& GHZ Inner Limit \\
		Moist greenhouse effect		&  3  &  0.95    &  0.97		& 0.993	& ... \\
		Earth-equivalent position	&  0  &  0.993   &  $\equiv$1 	& $\equiv$1	& ... \\
		First CO$_2$ condensation	&  4  &  1.37    &  1.40		& ... 	& ... \\
		Maximum greenhouse effect	&  5  &  1.67    &  1.71		& 1.676	& GHZ Outer Limit \\
		Early Mars				&  6  &  1.77    &  1.81		& 1.768	& RVEM Outer Limit \\
\enddata
\tablecomments{
This table depicts the various values of $s_\ell$ (see Eq.~1), as previously obtained in the literature.
Here Kas93 denotes the work by \cite{kas93}, and Kop1314 denotes the combined work by \cite{kop13} and
\cite{kop14}.
}
\end{deluxetable}

%+++++++++++++++++++++++++++++++++++++++++++++++++++++++++++++++++++++++

\clearpage

%%% *** Table 2
%%%%%%%%%%%%%%%%%%%%%%%%%%%%%%%%%%%%%%%%%%%%%%%%%%%%%%%%%%%%%%%%%%%%%%%%%
\begin{deluxetable}{ccccc}
\tablecaption{Stellar Parameters}
\tablewidth{0pt}
\tablehead{
$M_{*}$ & Spectral Type & $T_{*}$ & $R_{*}$ & $L_{*}$ \\
\noalign{\smallskip}
\hline
\noalign{\smallskip}
($M_{\odot}$) & ... & (K) & ($R_{\odot}$) & ($L_{\odot}$)
}
\startdata
		1.25 & $\sim$F6V & 6257 & 1.253 & 2.154    \\
		1.00 & $\sim$G2V & 5780 & 1.000 & 1.0000   \\
		0.75 & $\sim$K2V & 5104 & 0.766 & 0.3568   \\
		0.50 & $\sim$M0V & 3664 & 0.472 & 0.03593  \\
\enddata
\tablecomments{Adopted from Paper~I and II.
}
\end{deluxetable}

%+++++++++++++++++++++++++++++++++++++++++++++++++++++++++++++++++++++++

\clearpage

%%% *** Table 3
%%%%%%%%%%%%%%%%%%%%%%%%%%%%%%%%%%%%%%%%%%%%%%%%%%%%%%%%%%%%%%%%%%%%%%%%%
\begin{deluxetable}{lcccc}
\tablecaption{BIC Values for $a_{\rm bin}$ versus $e_b$ Fitting}
\tablewidth{0pt}
\tablehead{BIC & Linear & Quadratic & Cubic & Quartic}
\startdata
P$-$GHZ	& $-$287.28   	& $-$418.18  	& $-$538.74		& $-$627.83	\\
P$-$RVEM   	& $-$295.91  	& $-$432.93		& $-$564.12		& $-$642.10	\\
S$-$GHZ	& $-$7915.7	 	& $-$12327		& $-$16347		& $-$19794	\\
S$-$RVEM   	& $-$7422.6  	& $-$11599		& $-$15395		& $-$18840	\\	
\enddata
\tablecomments{The case of $M_{1} = M_{2} = 1.0~M_{\odot}$ is given as an example 
for the determination of the $a_{\rm bin}$ versus $e_{b}$ fitting.  For all S-type results,
the logarithm of $a_{\rm bin}$ is applied.  The mean absolute percentage error (MAPE) is
calculated as well, and a 2\% threshold is used.  The BICs are found to decrease as the orders of the equations
increase from 1 to 3 for all cases indicating that it is acceptable to have cubic equations.
Lowest order equations satisfy the MAPE requirement are chosen for less complexity.}
\end{deluxetable}

%+++++++++++++++++++++++++++++++++++++++++++++++++++++++++++++++++++++++

\clearpage

%%% *** Table 4
%%%%%%%%%%%%%%%%%%%%%%%%%%%%%%%%%%%%%%%%%%%%%%%%%%%%%%%%%%%%%%%%%%%%%%%%%
\begin{deluxetable}{lcrrrr}
\tablecaption{Fitting Coefficients}
\tablewidth{0pt}
\tablehead{Model & Coefficient & Case 1 & Case 2 & Case 3 & Case 4
}
\startdata
P-GHZ		& $\alpha_{0}$	& 0.215	& 0.805	& 0.805	& 1.185 \\
...	 	& $\alpha_{1}$	& $-$0.205	& $-$0.933	& $-$0.933	& $-$1.302 \\
...	 	& $\alpha_{2}$ 	& 0.123	& 0.716	& 0.716	& 0.949 \\
P-RVEM 	& $\alpha_{0}$ 	& 0.230	& 0.850	& 0.850	& 1.253 \\
...	 	& $\alpha_{1}$ 	& $-$0.228	& $-$0.957	& $-$0.957	& $-$1.371 \\
...	 	& $\alpha_{2}$ 	& 0.146	& 0.695	& 0.695	& 0.992 \\
\noalign{\smallskip}
\hline
\noalign{\smallskip}
S-GHZ	 	& $\beta_{0}$  	& $-$0.328	& 0.994	& 0.994	& 1.442 \\
...	 	& $\beta_{1}$  	& 2.169	& 2.088	& 2.088	& 1.601 \\
...	 	& $\beta_{2}$  	& $-$3.046	& $-$2.497	& $-$2.497	& $-$0.956 \\
...	 	& $\beta_{3}$  	& 4.346	& 3.901	& 3.901	& 2.506 \\
S-RVEM 	& $\beta_{0}$  	& $-$0.566	& 0.746	& 0.746	& 1.19 \\
...	 	& $\beta_{1}$  	& 2.177	& 2.256	& 2.256	& 1.844 \\
...	 	& $\beta_{2}$  	& $-$3.065	& $-$3.024	& $-$3.024	& $-$1.787 \\
...	 	& $\beta_{3}$  	& 4.359	& 4.349	& 4.349	& 3.275 \\
\enddata
\tablecomments{
Case~1: $M_{1} = M_{2} =  0.50~M_{\odot}$;
Case~2: $M_{1} = 1.00~M_{\odot}$, $M_{2} = 0.50~M_{\odot}$;
Case~3: $M_{1} = M_{2} = 1.00~M_{\odot}$;
Case~4: $M_{1} = 1.25~M _{\odot}$, $M_{2} = 0.75~M_{\odot}$.
}
\end{deluxetable}

%+++++++++++++++++++++++++++++++++++++++++++++++++++++++++++++++++++++++

\clearpage

%%% *** Table 5
%%%%%%%%%%%%%%%%%%%%%%%%%%%%%%%%%%%%%%%%%%%%%%%%%%%%%%%%%%%%%%%%%%%%%%%%%
\begin{landscape}
\begin{deluxetable}{llcccccc}
\tabletypesize{\scriptsize}
\tablecaption{BIC Values for Mass Fitting}
\tablewidth{0pt}
\tablehead{
Model & BIC & Linear to $M_1$ and $M_2$ & Adding $M_1^2$ and $M_2^2$ & Adding ${M_1}{M_2}$ and $M_1^2$ & Adding ${M_1}{M_2}$ and $M_2^2$ 
                       & Adding $M_1^2$ and $M_1^3$ & Adding $M_2^2$ and $M_2^3$
}
\startdata
P$-$GHZ 	& Constant			& $-$80.90		& $-$75.94		& $-$76.06		& $-$75.21		& {\bf $-$81.84}	& $-$75.94 \\
...    		& Coef. of $e_b$ term	& {\bf $-$71.59}	& $-$67.34		& $-$68.70		& $-$68.32		& $-$70.66		& $-$67.67 \\
...		& Coef. of $e_b^2$ term 	& {\bf $-$74.75}	& $-$69.81		& $-$71.19		& $-$70.77		& $-$74.42		& $-$69.65 \\
P$-$RVEM 	& Constant			& {\bf $-$78.20}	& $-$71.14		& $-$75.24		& $-$75.76		& $-$76.26		& $-$74.80 \\
...    		& Coef. of $e_b$ term	& $-$68.41		& $-$66.79		& {\bf $-$68.86}	& $-$64.14		& $-$65.96		& $-$63.09 \\
...		& Coef. of $e_b^2$ term 	& $-$64.78		& {\bf $-$66.50}	& $-$63.89		& $-$65.04		& $-$61.58		& $-$63.44 \\
\noalign{\smallskip}
\hline
\noalign{\smallskip}
Total		& ...				&  {\bf $-$438.64}	& $-$417.51		& $-$423.94		& $-$419.25		& $-$430.73		& $-$414.59	\\
\noalign{\smallskip}
\hline
\noalign{\smallskip}
S$-$GHZ 	& Constant			& $-$31.68		& $-$48.70		& $-$49.40		& $-$34.20		& {\bf $-$66.87}	& $-$28.96 \\
...    		& Coef. of $e_b$ term	& $-$41.16		& $-$51.60		& $-$51.75		& $-$38.90		& {\bf $-$57.37}	& $-$37.18 \\
...		& Coef. of $e_b^2$ term 	& $-$18.43		& $-$30.32		& $-$30.15		& $-$17.85		& {\bf $-$35.97}	& $-$14.79 \\
...		& Coef. of $e_b^3$ term 	& $-$20.18 		& $-$34.76		& $-$34.62		& $-$20.22		& {\bf $-$39.84}	& $-$16.63 \\
S$-$RVEM 	& Constant			& $-$31.78		& $-$46.96		& $-$47.44		& $-$33.85		& {\bf $-$69.32}	& $-$28.97 \\
...    		& Coef. of $e_b$ term	& $-$36.27		& $-$53.25		& $-$54.17		& $-$38.71		& {\bf $-$65.40}	& $-$33.20 \\
...		& Coef. of $e_b^2$ term 	& $-$13.49		& $-$28.14		& $-$30.12		& $-$17.55		& {\bf $-$36.86}	& $-$10.94 \\
...		& Coef. of $e_b^3$ term 	& $-$16.11		& $-$30.85		& $-$32.98		& $-$20.42		& {\bf $-$39.04}	& $-$13.67 \\
\noalign{\smallskip}
\hline
\noalign{\smallskip}
Total		& ...				& $-$209.11		& $-$324.58		& $-$330.62		& $-$221.70		& {\bf $-$410.68}	& $-$184.34	 \\
\enddata
\tablecomments{
The coefficients from the $a_{\rm bin}$ versus $e_{b}$ fitting are further fitted based on an equation linear
in the stellar masses.  Additional terms are added by checking the BIC.  The smallest BIC in each line
is given in bold font. The total BIC values in each column are compared.  For P-type, adding nothing 
is preferred, whereas for S-type, adding $M_1^2$ and $M_1^3$ turns out to be the best choice.
}
\end{deluxetable}
\end{landscape}

%+++++++++++++++++++++++++++++++++++++++++++++++++++++++++++++++++++++++

\clearpage

%%% *** Table 6
%%%%%%%%%%%%%%%%%%%%%%%%%%%%%%%%%%%%%%%%%%%%%%%%%%%%%%%%%%%%%%%%%%%%%%%%%
\begin{deluxetable}{lcrrr}
\tablecaption{General Fitting Coefficients, P-Type}
\tablewidth{0pt}
\tablehead{
Model & Coefficient & $A_{i0}$ & $A_{i1}$ & $A_{i2}$
}
\startdata
P-GHZ		& $\alpha_{0}$	& $-$0.541	& 1.201	& 0.318 \\
...		& $\alpha_{1}$	& 0.541	& $-$1.504	& 0.012 \\
...		& $\alpha_{2}$	& $-$0.345	& 1.219	& $-$0.275 \\
P-RVEM	& $\alpha_{0}$	& $-$0.570	& 1.266	& 0.338 \\
...	 	& $\alpha_{1}$	& 0.578	& $-$1.553	& $-$0.036 \\
...	 	& $\alpha_{2}$	& $-$0.384	& 1.213	& $-$0.187 \\
\enddata
\end{deluxetable}

%+++++++++++++++++++++++++++++++++++++++++++++++++++++++++++++++++++++++

\clearpage

%%% *** Table 7
%%%%%%%%%%%%%%%%%%%%%%%%%%%%%%%%%%%%%%%%%%%%%%%%%%%%%%%%%%%%%%%%%%%%%%%%%
\begin{deluxetable}{lcrrrrr}
\tablecaption{General Fitting Coefficients, S-Type}
\tablewidth{0pt}
\tablehead{
Model & Coefficient & $B_{i0}$ & $B_{i1}$ & $B_{i2}$ & $B_{i3}$ & $B_{i4}$
}
\startdata
S-GHZ		& $\beta_{0}$	& $-$5.871	& 16.88	& 0.499	& $-$14.932	& 4.688 \\
...		& $\beta_{1}$	& $-$0.002	& 8.934	& $-$0.464	& $-$9.835	& 3.160 \\
...		& $\beta_{2}$	& 2.852	& $-$24.115	& 1.018	& 26.776	& $-$8.352 \\
...	 	& $\beta_{3}$	& $-$0.414	& 19.528	& $-$0.991	& $-$21.219	& 6.371 \\
S-RVEM 	& $\beta_{0}$	& $-$6.354	& 17.873	& 0.500	& $-$16.177	& 5.168 \\
...	 	& $\beta_{1}$	& 3.311	& $-$4.510	& $-$0.461	& 7.080	& $-$3.349 \\
...	 	& $\beta_{2}$	& $-$7.312	& 17.184	& 0.954	& $-$25.040	& 11.499 \\
...	 	& $\beta_{3}$	& 8.047	& $-$14.891	& $-$0.883	& 21.811	& $-$10.027 \\
\enddata
\end{deluxetable}

%+++++++++++++++++++++++++++++++++++++++++++++++++++++++++++++++++++++++

\clearpage

%%% *** Table 8
%%%%%%%%%%%%%%%%%%%%%%%%%%%%%%%%%%%%%%%%%%%%%%%%%%%%%%%%%%%%%%%%%%%%%%%%%
\begin{deluxetable}{ccccccccc}
\tablecaption{Errors of Fitting}
\tablewidth{0pt}
\tablehead{
$e_{b}$ & \multicolumn{4}{c}{$M_{1} = M_{2} = 1.00~M_{\odot}$} &
          \multicolumn{4}{c}{$M_{1} = 1.00~M_{\odot}, M_{2} = 0.50~M_{\odot}$} \\
\noalign{\smallskip}
\hline
\noalign{\smallskip}
... & P-GHZ & P-RVEM & S-GHZ & S-RVEM & P-GHZ & P-RVEM & S-GHZ & S-RVEM
}
\startdata
		0.0 & 0.53\% & 0.17\% & 4.96\% & 6.36\% & 0.58\% & 0.02\% & 2.71\% & 4.09\%  \\ 
		0.1 & 2.28\% & 2.21\% & 0.67\% & 0.44\% & 2.25\% & 2.20\% & 2.21\% & 2.46\%  \\ 
		0.2 & 2.66\% & 2.46\% & 0.11\% & 0.89\% & 2.27\% & 2.33\% & 3.26\% & 4.09\%  \\ 
		0.3 & 2.20\% & 2.04\% & 0.70\% & 0.09\% & 1.31\% & 1.38\% & 2.14\% & 2.80\%  \\ 
		0.4 & 1.59\% & 1.40\% & 1.59\% & 1.52\% & 0.33\% & 0.19\% & 0.66\% & 0.86\%  \\ 
		0.5 & 1.27\% & 1.11\% & 1.82\% & 2.20\% & 0.07\% & 0.40\% & 0.18\% & 0.20\%  \\ 
		0.6 & 1.59\% & 1.50\% & 0.98\% & 1.40\% & 1.20\% & 0.16\% & 1.13\% & 0.62\%  \\ 
		0.7 & 2.99\% & 2.98\% & 0.67\% & 0.08\% & 4.13\% & 2.32\% & 2.27\% & 2.59\%   \\ 
		0.8 & 5.78\% & 5.82\% & ...         & ...         & 9.26\% & 6.34\% & ...         & 0.92\%    \\
\enddata
\end{deluxetable}

%+++++++++++++++++++++++++++++++++++++++++++++++++++++++++++++++++++++++

\clearpage

%%% *** Table 9
%%%%%%%%%%%%%%%%%%%%%%%%%%%%%%%%%%%%%%%%%%%%%%%%%%%%%%%%%%%%%%%%%%%%%%%%%
\begin{deluxetable}{ccccccccc}
\tablecaption{Errors of Fitting, Continued}
\tablewidth{0pt}
\tablehead{
$e_{b}$ & \multicolumn{4}{c}{$M_{1} = 0.75~M_{\odot}, M_{2} = 0.50~M_{\odot}$} &
          \multicolumn{4}{c}{$M_{1} = M_{2} = 0.50~M_{\odot}$} \\
\noalign{\smallskip}
\hline
\noalign{\smallskip}
... & P-GHZ & P-RVEM & S-GHZ & S-RVEM & P-GHZ & P-RVEM & S-GHZ & S-RVEM
}
\startdata
		0.0 & 3.48\% & 3.47\% & 2.97\% & 3.36\% & 0.33\% & 0.00\% & 5.41\% & 6.08\%  \\ 
		0.1 & 1.87\% & 1.85\% & 3.77\% & 3.80\% & 1.82\% & 2.05\% & 0.58\% & 0.59\%  \\ 
		0.2 & 2.02\% & 1.95\% & 5.40\% & 5.52\% & 2.50\% & 2.74\% & 2.30\% & 2.26\%  \\ 
		0.3 & 2.96\% & 2.92\% & 4.04\% & 4.12\% & 2.37\% & 2.53\% & 1.21\% & 1.09\%  \\ 
		0.4 & 3.99\% & 4.14\% & 1.91\% & 2.12\% & 2.00\% & 1.94\% & 0.65\% & 0.75\%  \\ 
		0.5 & 4.38\% & 4.83\% & 0.87\% & 0.98\% & 1.80\% & 1.49\% & 1.60\% & 1.59\%  \\ 
		0.6 & 3.58\% & 4.52\% & 1.63\% & 1.63\% & 2.25\% & 1.47\% & 0.67\% & 0.65\%  \\ 
		0.7 & 1.29\% & 2.88\% & 3.43\% & 3.34\% & 3.60\% & 2.33\% & 1.24\% & 1.12\%  \\ 
		0.8 & 2.58\% & 0.45\% & 0.34\% & 0.46\% & 6.10\% & 4.56\% & 2.27\% & 2.86\%  \\
\enddata
\end{deluxetable}

%+++++++++++++++++++++++++++++++++++++++++++++++++++++++++++++++++++++++

\clearpage

%%% *** Table 10
%%%%%%%%%%%%%%%%%%%%%%%%%%%%%%%%%%%%%%%%%%%%%%%%%%%%%%%%%%%%%%%%%%%%%%%%%
\begin{deluxetable}{lcccc}
\tablecaption{Coefficient of Determination}
\tablewidth{0pt}
\tablehead{
Systems & P-GHZ & P-RVEM & S-GHZ & S-RVEM
}
\startdata
	$M_{1} = 1.25~M_{\odot}, M_{2} = 1.25~M_{\odot}$ & 0.9976 & 0.9975 & 0.9967 & 0.9969 \\
	$M_{1} = 1.25~M_{\odot}, M_{2} = 1.00~M_{\odot}$ & 0.9795 & 0.9758 & 0.9999 & 0.9997 \\
	$M_{1} = 1.25~M_{\odot}, M_{2} = 0.75~M_{\odot}$ & 0.9919 & 0.9929 & 0.9990 & 0.9991 \\
	$M_{1} = 1.25~M_{\odot}, M_{2} = 0.50~M_{\odot}$ & 0.9954 & 0.9940 & 0.9991 & 0.9990 \\
	$M_{1} = 1.00~M_{\odot}, M_{2} = 1.00~M_{\odot}$ & 0.9803 & 0.9820 & 0.9990 & 0.9984 \\
	$M_{1} = 1.00~M_{\odot}, M_{2} = 0.75~M_{\odot}$ & 0.9970 & 0.9963 & 0.9994 & 0.9988 \\
	$M_{1} = 1.00~M_{\odot}, M_{2} = 0.50~M_{\odot}$ & 0.9849 & 0.9853 & 0.9985 & 0.9978 \\
	$M_{1} = 0.75~M_{\odot}, M_{2} = 0.75~M_{\odot}$ & 0.9876 & 0.9861 & 0.9947 & 0.9954 \\
	$M_{1} = 0.75~M_{\odot}, M_{2} = 0.50~M_{\odot}$ & 0.9619 & 0.9632 & 0.9960 & 0.9962 \\
	$M_{1} = 0.50~M_{\odot}, M_{2} = 0.50~M_{\odot}$ & 0.9780 & 0.9834 & 0.9991 & 0.9991 \\
\enddata
\end{deluxetable}

%+++++++++++++++++++++++++++++++++++++++++++++++++++++++++++++++++++++++

\clearpage

%% *** Table 11
%%%%%%%%%%%%%%%%%%%%%%%%%%%%%%%%%%%%%%%%%%%%%%%%%%%%%%%%%%%%%%%%%%%%%%%%
\thispagestyle{empty} 
\begin{landscape}
\begin{deluxetable}{lccccl}
% \tabletypesize{\scriptsize}
\tablecaption{System Parameters}
\tablewidth{0pt}
\tablehead{
System & $M_{1}$       & $M_{2}$       & $a_{\rm bin}$  & $e_{b}$ & Reference \\
...    & ($M_{\odot}$) & ($M_{\odot}$) & (au) & ...     &  ...
}
\startdata
	Kepler-34	& $1.0479^{+0.0033}_{-0.0030}$ & 1.0208$\pm$0.0022	        & $0.22882^{+0.00019}_{-0.00018}$	& $0.52087^{+0.00052}_{-0.00055}$ & \cite{wel12} \\
\noalign{\smallskip}
	Kepler-35	& $0.8877^{+0.0051}_{-0.0053}$ & $0.8094^{+0.0042}_{-0.0045}$ & $0.17617^{+0.00029}_{-0.00030}$	& $0.1421^{+0.0014}_{-0.0015}$    & \cite{wel12} \\
\noalign{\smallskip}
	Kepler-413	& $0.820^{+0.015}_{-0.014}$	 & $0.5423^{+0.0081}_{-0.0073}$ & $0.10148^{+0.00057}_{-0.00052}$	& $0.0365^{+0.0023}_{-0.0021}$    & \cite{kos14} \\
\noalign{\smallskip}
	Kepler-1647	& 1.2207$\pm$0.0112	       & 0.9678$\pm$0.0039	        & 0.1276$\pm$0.0002	            & 0.1602$\pm$0.0004	          & \cite{kos16} \\
\noalign{\smallskip}
	TrES-2	& 1.05	                   & 0.67	                    & 232$\pm$12	                  & ...	                            & \cite{dae09} \\
\noalign{\smallskip}
	KOI-1257	& 0.99$\pm$0.05	             & 0.70$\pm$0.07	              & 5.3$\pm$1.3	                  & $0.31^{+0.37}_{-0.21}$	    & \cite{san14} \\
\enddata
\end{deluxetable}
\end{landscape}


\begin{thebibliography}{}

\bibitem[Bazs{\'o} et al.(2017)]{baz17}
Bazs{\'o}, {\'A}., Pilat-Lohinger, E., Eggl, S., Funk, B., Bancelin, D.,
\& Rau, G. 2017, \mnras, 466, 1555

\bibitem[Chabrier(2003)]{cha03}
Chabrier, G. 2003, \pasp, 115, 763

\bibitem[Cuntz(2014)]{cun14}
Cuntz, M. 2014, \apj, 780, 14~[Paper I]

\bibitem[Cuntz(2015)]{cun15}
Cuntz, M. 2015, \apj, 798, 101~[Paper II]

\bibitem[Cuntz \& Bruntz(2014)]{cunb14}
Cuntz, M., \& Bruntz, R. 2014, in Cool Stars, Stellar Systems, and the Sun:
18th Cambridge Workshop, ed. G. van Belle \& H. Harris (Flagstaff: Lowell
Observatory), p. 831

\bibitem[Daemgen et al.(2009)]{dae09}
Daemgen, S., Hormuth, F., Brandner, W., Bergfors, C., Janson, M., Hippler, S.,
\& Henning, T. 2009, \aap, 498, 567

\bibitem[Duquennoy \& Mayor(1991)]{duq91}
Duquennoy, A., \& Mayor, M. 1991, \aap, 248, 485

\bibitem[Dvorak(1982)]{dvo82}
Dvorak, R. 1982, OAWMN, 191, 423

\bibitem[Dvorak(1986)]{dvo86}
Dvorak, R. 1986, \aap, 167, 379

\bibitem[Eggenberger et al.(2004)]{egg04}
Eggenberger, A., Udry, S., \& Mayor, M. 2004, \aap, 417, 353

\bibitem[Eggl et al.(2012)]{egg12}
Eggl, S., Pilat-Lohinger, E., Georgakarakos, N., Gyergyovits, M., \&
Funk, B. 2012, \apj, 752, 74

\bibitem[Eggl et al.(2013)]{egg13}
Eggl, S., Pilat-Lohinger, E., Funk, B., Georgakarakos, N., \&
Haghighipour, N. 2013, \mnras, 428, 3104

\bibitem[Gray(2005)]{gra05}
Gray, D. F. 2005, The Observation and Analysis of Stellar Photospheres,
2nd edn., Cambridge University Press, Cambridge

\bibitem[Haghighipour(2008)]{hag08}
Haghighipour, N. 2008, in Exoplanets, Springer Praxis Books, Chichester, UK,
p. 223

\bibitem[Haghighipour \& Kaltenegger(2013)]{hag13}
Haghighipour, N., \& Kaltenegger, L. 2013, \apj, 777, 166

\bibitem[Holman \& Wiegert(1999)]{hol99}
Holman, M. J., \& Wiegert, P. A. 1999, \aj, 117, 621

\bibitem[Kaltenegger \& Haghighipour(2013)]{kal13}
Kaltenegger, L., \& Haghighipour, N. 2013, \apj, 777, 165

\bibitem[Kane \& Hinkel(2013)]{kan13}
Kane, S. R., \& Hinkel, N. R. 2013, \apj, 762, 7

\bibitem[Kasting et al.(1993)]{kas93}
Kasting, J. F., Whitmire, D. P., \& Reynolds, R. T. 1993,
Icarus, 101, 108

\bibitem[Kopparapu et al.(2013)]{kop13}
Kopparapu, R. K., Ramirez, R., Kasting, J. F., et al. 2013, \apj, 765, 131;
Erratum 770, 82

\bibitem[Kopparapu et al.(2014)]{kop14}
Kopparapu, R. K., Ramirez, R. M., SchottelKotte, J., et al. 2014,
\apj, 787, L29

\bibitem[Kostov et al.(2014)]{kos14}
Kostov, V. B., McCullough, P. R., Carter, J. A., et al. 2014,
\apj, 784, 14; Erratum 787, 93

\bibitem[Kostov et al.(2016)]{kos16}
Kostov, V. B., Orosz, J. A., Welsh, W. F., et al. 2016,
\apj, 827, 86

\bibitem[Kroupa(2001)]{kro01}
Kroupa, P. 2001, \mnras, 322, 231

\bibitem[Kroupa(2002)]{kro02}
Kroupa, P. 2002, Science, 295, 82

\bibitem[Ma et al.(2016)]{ma16}
Ma, B., Ge, J., Wolszczan, A., et al. 2016, \aj, 152, 112

\bibitem[Mann et al.(2013)]{man13}
Mann, A. W., Gaidos, E., \& Ansdell, M. 2013, \apj, 779, 188

\bibitem[M\"uller \& Haghighipour(2014)]{mul14}
M\"uller, T. W. A., \& Haghighipour, N. 2014, \apj, 782, 26

\bibitem[Ortiz et al.(2015)]{ort15}
Ortiz, M., Gandolfi, D., Reffert, S., et al. 2015, \aap, 573, L6

\bibitem[Patience et al.(2002)]{pat02}
Patience, J., White, R. J., Ghez, A. M., et al. 2002, \apj, 581, 654

\bibitem[Pilat-Lohinger \& Dvorak(2002)]{pil02}
Pilat-Lohinger, E., \& Dvorak, R. 2002, CMDA, 82, 143

\bibitem[Rabl \& Dvorak(1988)]{rab88}
Rabl, G., \& Dvorak, R. 1988, \aap, 191, 385

\bibitem[Raghavan et al.(2006)]{rag06}
Raghavan, D., Henry, T. J., Mason, B. D., et al. 2006, \apj, 646, 523

\bibitem[Raghavan et al.(2010)]{rag10}
Raghavan, D., McAlister, H. A., Henry, T. J., et al. 2010, \apjs, 190, 1

\bibitem[Roell et al.(2012)]{roe12}
Roell, T., Neuh\"auser, R., Seifahrt, A., \& Mugrauer, M. 2012,
\aap, 542, A92

\bibitem[Santerne et al.(2014)]{san14}
Santerne, A., H{\'e}brard, G., Deleuil, M., et al. 2014, \aap, A571, 37

\bibitem[Underwood et al.(2003)]{und03}
Underwood, D. R., Jones, B. W., \& Sleep, P. N. 2003, IJAsB, 2, 289

\bibitem[Wang \& Cuntz(2016)]{wan16}
Wang, Zh., \& Cuntz, M. 2016, in The 19th Cambridge Workshop on Cool Stars,
Stellar Systems, and the Sun,
ed. G. A. Feiden, 7, zenodo, doi:10.5281/zenodo.154425

\bibitem[Welsh et al.(2012)]{wel12}
Welsh, W. F., Orosz, J. A., Carter, J. A., et al. 2012, \nat, 481, 475

\bibitem[Welsh et al.(2015)]{wel15}
Welsh, W. F., Orosz, J. A.; Short, D. R., et al. 2015, \apj, 809, 26

\bibitem[Yi et al.(2001)]{yi01}
Yi, S., Demarque, P., Kim, Y.-C., Lee, Y.-W., Ree, C. H., Lejeune, T., \& Barnes, S.
2001, \apjs, 136, 417

\bibitem[Zuluaga et al.(2016)]{zul16}
Zuluaga, J. I., Mason, P. A., \& Cuartas-Restrepo, P. A. 2016,
\apj, 818, 160

\end{thebibliography}
\end{document}